\documentclass[aps,twocolumn,groupedaddress,nofootinbib,longbibliography,notitlepage,floatfix]{revtex4-2}
\usepackage{graphicx,outlines,subfigure}
\usepackage{physics,graphicx,diagbox,markdown}
\usepackage{tabularx}
\usepackage{amsmath,mathtools}
\usepackage{amstext,cuted}
\usepackage{amssymb}
\usepackage{xfrac}
\usepackage[colorlinks,citecolor=magenta]{hyperref}
\usepackage{graphicx,makecell}
\usepackage{amsmath}
\usepackage{amstext}
\usepackage{amssymb}
\usepackage{longtable,booktabs}
\usepackage{hyperref}
\usepackage{url}
\usepackage{subfigure}
\usepackage{dsfont}
\usepackage{booktabs}
\usepackage{amsbsy}
\usepackage{dcolumn}
\usepackage{amsthm}
\usepackage{times}
\usepackage{bm}
\usepackage{esint}
\usepackage{multirow}
\usepackage{hyperref}
\usepackage{cleveref}
\usepackage{amsmath}
\usepackage{amssymb}
\usepackage{mathrsfs}
\usepackage{amsbsy}
\usepackage{dcolumn}
\usepackage{bm}
\usepackage{multirow}
\usepackage{color}
\usepackage{extarrows}
\usepackage{datetime}
\usepackage{comment}
\usepackage[super]{nth}
\usepackage{tikz}
\usetikzlibrary{arrows.meta}
\usetikzlibrary{mindmap}
\usetikzlibrary{positioning}
\hypersetup{
	colorlinks=true,
	linkcolor=blue,
	filecolor=blue,
	urlcolor=blue,
}

\usepackage{datetime}
\usepackage{comment}
\usepackage{lineno}
\usepackage{xcolor}
\usepackage{scalerel}
\usepackage{tikz}
\usetikzlibrary{svg.path}

\definecolor{orcidlogocol}{HTML}{A6CE39}
\tikzset{
  orcidlogo/.pic={
    \fill[orcidlogocol] svg{M256,128c0,70.7-57.3,128-128,128C57.3,256,0,198.7,0,128C0,57.3,57.3,0,128,0C198.7,0,256,57.3,256,128z};
    \fill[white] svg{M86.3,186.2H70.9V79.1h15.4v48.4V186.2z}
                 svg{M108.9,79.1h41.6c39.6,0,57,28.3,57,53.6c0,27.5-21.5,53.6-56.8,53.6h-41.8V79.1z M124.3,172.4h24.5c34.9,0,42.9-26.5,42.9-39.7c0-21.5-13.7-39.7-43.7-39.7h-23.7V172.4z}
                 svg{M88.7,56.8c0,5.5-4.5,10.1-10.1,10.1c-5.6,0-10.1-4.6-10.1-10.1c0-5.6,4.5-10.1,10.1-10.1C84.2,46.7,88.7,51.3,88.7,56.8z};
  }
}

\newcommand\orcidicon[1]{\href{https://orcid.org/#1}{\mbox{\scalerel*{
\begin{tikzpicture}[yscale=-1,transform shape]
\pic{orcidlogo};
\end{tikzpicture}
}{|}}}}

 \usepackage{amsmath}
\usepackage{amssymb}
\usepackage{graphicx}
\usepackage{dcolumn}
\usepackage{bm}
\usepackage{subfigure}
\usepackage{slashed}
\usepackage{color}
\usepackage{epsf}

\usepackage{amsfonts}
\usepackage{dsfont}
\usepackage{wrapfig}
\usepackage{hyperref}
\usepackage{mathrsfs}
\usepackage{tikz}

\usepackage{braket}
\usepackage{indentfirst}
\usepackage{cases}
\usepackage{diagbox}
\usepackage{multirow}
\usepackage{makecell}
\usepackage{blkarray}
\usepackage{longtable}
\usepackage{extarrows}

\newcommand{\mT}{T}

\newcommand{\p}{\partial}

\newcommand{\ZZ}{\mathbb{Z}}

\newcommand{\lb}{\left(}
\newcommand{\rb}{\right)}
\newcommand{\bs}[1]{\mathbf{#1}}      
\renewcommand{\bf}{\mathbf}

\begin{document}

 \preprint{APS/123-QED}
\title{Three-dimensional fracton topological orders with boundary Toeplitz braiding}

\author{Bo-Xi Li}
\thanks{These authors contributed equally to this work.}
\affiliation{Guangdong Provincial Key Laboratory of Magnetoelectric Physics and Devices, State Key Laboratory of Optoelectronic Materials and Technologies,
and School of Physics, Sun Yat-sen University, Guangzhou, 510275, China}

\author{Yao Zhou}
\thanks{These authors contributed equally to this work.}
\affiliation{Guangdong Provincial Key Laboratory of Magnetoelectric Physics and Devices, State Key Laboratory of Optoelectronic Materials and Technologies,
and School of Physics, Sun Yat-sen University, Guangzhou, 510275, China}

\author{Peng Ye\orcidicon{0000-0002-6251-677X}}

\email{yepeng5@mail.sysu.edu.cn}

\affiliation{Guangdong Provincial Key Laboratory of Magnetoelectric Physics and Devices, State Key Laboratory of Optoelectronic Materials and Technologies,
and School of Physics, Sun Yat-sen University, Guangzhou, 510275, China}

\date{\today}

\begin{abstract}
In this paper, we theoretically study a class of 3D non-liquid states that show exotic boundary phenomena in the thermodynamical limit.   More concretely, we focus on a class of 3D fracton topological orders formed via  stacking 2D twisted \(\mathbb{Z}_N\) topologically ordered layers along   \(z\)-direction. Nearby layers are coupled while maintaining translation symmetry along  \(z\) direction. The effective field theory is given by the infinite-component Chern-Simons (iCS) field theory, with an integer-valued symmetric block-tridiagonal Toeplitz \(K\)-matrix whose size is thermodynamically large. With open boundary conditions (OBC) along \(z\), certain choice of \(K\)-matrices exhibits exotic boundary ``Toeplitz braiding'',  where the mutual braiding phase angle between two anyons  at opposite boundaries oscillates and remains non-zero in the thermodynamic limit.  In contrast, in trivial case,  the mutual braiding phase angle decays exponentially to zero in the thermodynamical limit.  As a necessary condition, this phenomenon requires the existence of boundary zero modes in the \(K\)-matrix spectrum under OBC.  We categorize nontrivial \(K\)-matrices into two distinct  types. Each type-I  possesses two boundary zero modes, whereas each type-II possesses only one boundary zero mode. Interestingly,   the  integer-valued Hamiltonian matrix of the familiar 1D  ``Su-Schrieffer-Heeger model'' can be used as a non-trivial $K$ matrix. Importantly, since  large-gauge-invariance ensures integer quantized \(K\)-matrix entries, global symmetries are not needed to protect these zero modes. We also  present numerical simulation as well as finite size scaling, further confirming the above analytical results. Symmetry fractionalization in iCS field theory is also briefly discussed. Motivated by the present field-theoretical work, it will be interesting to construct 3D lattice models for demonstrating Toeplitz braiding, which is left to future investigation.
 \end{abstract}
 
\maketitle


\section{Introduction}

Recently, research on topologically-ordered non-liquid states~\cite{2015arXiv150802595Z} has flourished, notably \textit{fracton topological order} (FTO)~\cite{nandkishore2019fractons}. These strongly-correlated systems with long-range entanglement have point-like excitations with limited mobility, unlike conventional intrinsic topological orders where excitations move freely without extra energy cost. FTOs' sensitivity to lattice details prevents a conventional continuum limit, setting them apart from liquid states. Exactly solvable models like the X-cube model~\cite{vijay2016fracton} and Haah’s code~\cite{haah2011local} reveal these properties. A large class of FTO models, including the X-cube model, display a foliation structure, showing that some (3+1)D\footnote{In this paper, ``$n+1$D" refers to $(n+1)$-dimensional spacetime, while ``$n$D" refers to $n$ spatial dimensions.} FTOs can be seen as stacks of coupled (2+1)D topological orders. From the entanglement perspective, extracting layers of (2+1)D topological orders resembles ``coarse graining" within the entanglement renormalization group framework. Here, the X-cube model is the fixed point of entanglement renormalization~\cite{shirley2018fracton}, and the entanglement renormalization group provides a hierarchical structure~\cite{li2023hierarchy} for various topologically-ordered non-liquid states~\cite{li2020fracton,li2021fracton}.

As  the analysis of lattice models indicates that many FTOs can be constructed by properly stacking and coupling lower-dimensional topologically-ordered liquid states. This prompts the question of what the field-theoretic counterpart is for the process of stacking topological orders. Along this line of thinking,  one solution is to take full use of so-called infinite-component Chern-Simons theory (iCS), which is  a very useful and powerful (3+1)D field theory recently studied in
Refs.~\cite{ma2022fractonic,chen2022gapless,chen2023ground,wu2023transition,sullivan2021weak,sullivan2021planar,levin2009gapless}.  Let us elaborate more. 

It is known that the effective theories of (2+1)D Abelian topological orders are $K$-matrix Chern-Simons theories~\cite{wen1992classification,wen2004quantum}, which provide a framework to understand a large class of topologically-ordered states, including multilayer fractional quantum Hall effects, spin liquids, and others. They also offer a systematic platform to classify symmetry-enriched topological (SET) and symmetry-protected topological (SPT) phases in (2+1)D, see, e.g., Refs.~\cite{lu2012theory,lu2016classification,YW12,Ye14b,hung2013kmatrix,PhysRevB.93.115136,meng2014topological}.  In the following, we assume all these (2+1)D topological orders reside in the spacetime with coordinates  parameterized by $x$, $y$, $t$. So, the wavefunctions of these states are defined on an $xy$ plane.
Next, one can stack each (2+1)D theory one by one along $z$ direction in order to form a multilayer quantum state. If the layer number tends to infinity, one   ends up with an authentic  (3+1)D quantum state. For example, one can start with the Laughlin state described by the Chern-Simons theory $\mathcal{L} = \frac{m}{4\pi} a_\mu \partial_\nu a_\lambda \epsilon^{\mu\nu\lambda}$, where $a_\mu$ is a $U(1)$ gauge field and $m \in \mathbb{Z}$. By stacking the Laughlin states and coupling the nearby layers, we obtain the effective theory for (3+1)D fractional quantum Hall states. The transport properties of the lateral surfaces of these states have been widely studied both theoretically and experimentally~\cite{balents1996chiral,kane1995impurity,naud2000fractional,naud2001notes,druist1998observation,qiu1989phases}. The resulting low-energy effective theory of a multilayer fractional quantum Hall state is a multi-component Chern-Simons (mCS) theory, described by the Lagrangian $\mathcal{L} = \sum_{I=1}^N \frac{m}{4\pi} a_\mu^I \partial_\nu a_\lambda^I \epsilon^{\mu\nu\lambda}$, where $N$ refers to the number of layers. In this context, the superscript $I$, originally defined as the layer index,  represents the location of the gauge field along $z$ direction. If $N$ is pushed to infinity, the resulting theory is an aforementioned infinite-component Chern-Simons, i.e., iCS field theory.  

\begin{figure}[htbp!]
    \centering
\includegraphics[width=8cm]{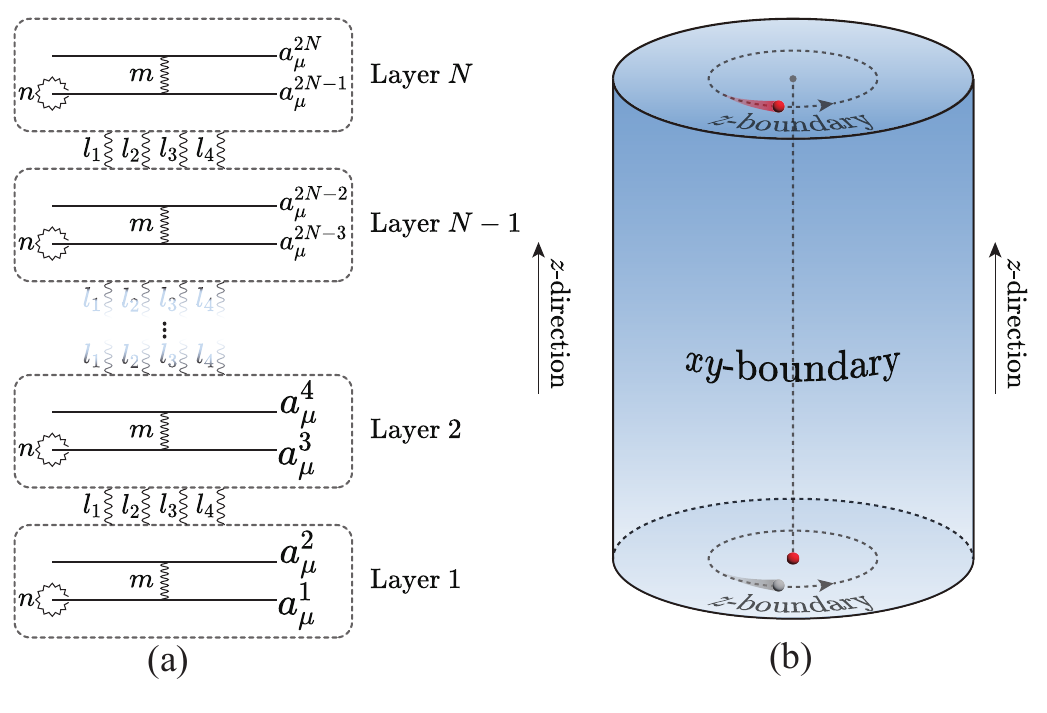}
\caption{(a) is an illustrative diagram of stacking $\ZZ_m$ twisted topological orders.  After stacking these $\ZZ_m$ twisted topological orders, we obtain a 3-dimensional theory as illustrated in (b). In the following discussion, open boundary condition is applied in $z$-direction. In our discussion, we focus on the asymptotic behaviour as the number of layer $N$ goes to infinity. (b) is also an illustrative diagram of braiding process between excitations residing at distinct \textbf{$\boldsymbol z$-boundaries}. The topological excitations are plotted as red dots in (b). If the $K$ matrix of the iCS field theory has boundary zero mode, together with the condition that the upper-right and lower-left entries of $K^{-1}$ are non-integers, two topological excitations residing at distinct $z$-boundaries can be detected mutually via mutual braiding (i.e., Toeplitz braiding) in the thermodynamical limit.  
}\label{stackingZNtopo}
\end{figure}

In this paper, we focus on    iCS field theories in which each layer is  a \textit{twisted $\mathbb{Z}_m$ topological order} described by a two-component  Chern-Simons theory $\frac{1}{4\pi} \sum^{2}_{I,J=1} \mathcal{K}_{IJ} a_\mu^I \partial_\nu a_\lambda^J \epsilon^{\mu\nu\lambda}$. Here, $\mathcal{K} = \left(\begin{smallmatrix} n & m \\ m & 0 \end{smallmatrix}\right)$ with $m,n\in\mathbb{Z}$.   $a_\mu^I$ is a compact $U(1)$  gauge field restricted on the $I$-th layer, meaning that  $I$ may serve as the coordinate along $z$ direction (assuming that each layer is on $xy$ plane).   
After stacking these (2+1)D twisted topological orders and coupling the nearby layers by additional Chern-Simons terms (see Fig.~\ref{stackingZNtopo}a), we can find that the $K$ matrix of the resulting iCS field theory is  a block-tridiagonal Toeplitz matrix (to be shown in Eq.~(\ref{ktridiagonal}) in Sec.~\ref{sectionII}).\footnote{In the following text, the term ``$K$ matrix" primarily refers to the Toeplitz $K$ matrix, as represented by Eq.~(\ref{ktridiagonal}), unless explicitly stated otherwise.} Locality of the resulting 3D theory requires that $K_{IJ}$ must decay no slower than exponential decay as $|I-J|$ increases. For the sake of simplicity, we only consider the coupling between   nearest (2+1)D topological orders.

Mathematically, Toeplitz matrices are defined as matrices possessing translation symmetry along each diagonal. They find wide applications in both physics and mathematics~\cite{lee2015free,meurant1992,hirschman1967matrix,ekstrom2018eigenvalues,wax1983efficient,gutierrez2012block,heinig1984algebraic}. When   individual entries of a Toeplitz matrix are replaced by matrix blocks, the resulting matrix is referred to as a block Toeplitz matrix. Interestingly, the formalism of the block-tridiagonal Toeplitz matrix immediately reminds us of   a  single-particle Hamiltonian of a 1D tight-binding  lattice fermion model. Research on topological insulators (TIs) and topological superconductors (TSCs) has shed light on some Toeplitz matrices possessing eigenvectors localized at boundaries~\cite{su1979solitons,qi2011topological,guo2016brief,kitaev2001unpaired,shen2012topological,asboth2016short} when open boundary condition (OBC) is imposed. These eigenvectors have eigenvalues that decrease exponentially to zero as the size of the matrix increases, which is well-known as \textit{boundary zero modes} in the  thermodynamical limit.  {We must stress that, such an analogy is made only on the level of similar mathematics.}  It is important to note that  $K$ matrix here is \textit{not} a Hamiltonian, and all matrix elements  here are strictly quantized to integers due to large gauge invariance.
As is known from the study of TIs and TSCs, boundary zero modes contain rich topological physics. Inspired by research in TIs and TSCs, we wonder  {whether and how the potential presence of  boundary zero modes of the matrix $K$ reshape the low-lying energy physics of (3+1)D fracton topological order described by iCS field theory.}  Especially,   in the presence of boundary zero modes of $K$ matrices, do there exist nontrivial properties on $z$-boundaries (i.e.,  two boundaries along $z$ direction) of iCS field theory? 

 In this paper, we investigate these questions through  certain iCS field theories under  OBC in the $z$-direction (denoted as $z$-OBC). Specifically, we couple twisted $\mathbb{Z}_m$ topological orders layer-by-layer, which leads to  $K$ matrices that take the form of block Toeplitz matrices.  It should be noted that, again, the ``boundary zero modes'' of Toeplitz $K$ matrices discussed here are not physical edge modes of TIs. Instead, they are, in a pure mathematical sense, eigenvalues of dimensionless $K$ matrices with OBC, which decrease exponentially to zero as the size of the $K$ matrix, i.e., the number of layers $N$, increases from finite to infinite.  Due to the presence of boundary zero modes in the $K$ matrix, we observe that the braiding statistics between topological excitations, even when widely separated along the $z$ direction, may be nontrivial, except for some very special cases. As an extreme example,  Fig.~\ref{stackingZNtopo}(b) illustrates the  braiding process between topological excitations residing at distinct $z$-boundaries, which also clarifies how  the braiding of two excitations is practically performed. In this paper,  this mutual braiding statistics is termed ``\textbf{Toeplitz braiding}'' in order to highlight the nontrivial mathematical properties of Toeplitz matrices.  

More analytically, we find that the presence of Toeplitz braiding necessitates not only the existence of boundary zero modes of $K$ matrices but also non-integer values of the upper-right and lower-left components of $K^{-1}$ (to be shown in Fig.~\ref{conditionToeplitz}(b) in the main text). We classify $K$ matrices that potentially support Toeplitz braiding into two distinct types and numerically verify how the braiding phase angle constantly oscillates as the number of layers along the $z$ direction tends to infinity. To make a clear  comparison, we also numerically compute a trivial case, where the mutual braiding phase angle exponentially decays to zero in the thermodynamical limit. By taking full advantage of  the properties of Toeplitz matrices and making a tight connection to the calculation of braiding statistics of Chern-Simons gauge theory,   our work  presented here   illuminates the exotic boundary topological physics of 3D strongly-correlated states,  advancing  the exploration of topologically ordered non-liquids including  fracton topological orders. Several interesting future directions are straightforward.

The paper is organized as follows. In Sec.~\ref{sectionII}, we carry on an analytic analysis on a concrete iCS field theory with Toeplitz braiding, whose $K$ matrix is, in its mathematical form,  equivalent to  the Hamiltonian of Su-Schrieffer-Heeger model~\cite{su1979solitons}.
In Sec.~\ref{sectionIII}, we propose a general framework for sorting $K$ matrices with boundary zero modes. Within this framework, we identify two distinct types of $K$ matrices that possess boundary zero modes, providing them as illustrative examples. The results are shown in Table \ref{icstoeplitz}.
To demonstrate the correspondence between boundary zero modes of $K$ matrices and nonlocal braiding statistics in $z$-direction in these models, we conduct numerical computations in Sec.~\ref{sectionIV}. Numerical computations on symmetry fractionalization 
of iCS field theories with Toeplitz braiding are also briefly discussed in Sec.~\ref{sectionIV}. This work is concluded in Sec.~\ref{sectionV}. 
\begin{table*}[htbp]
\caption{A class of Toeplitz $K$ matrices (Eq.~(\ref{ktridiagonal})) with boundary zero modes, where the block $A=\begin{pmatrix}
	n & m \\ m& 0
\end{pmatrix}$ and the block $B=\begin{pmatrix}
	l_1 & l_2 \\ l_3& l_4
\end{pmatrix}$. Type-I $K$ matrix possesses 2 boundary zero modes, whereas Type-II $K$ matrix possesses 1 boundary zero mode. Considering the physical meaning of block A, we require $m\neq 0$. It should be noted that the ``requirement" column is not the sufficient condition for \textit{Toeplitz braiding}.}
\label{icstoeplitz}
\begin{ruledtabular}
\begin{tabular}{ccc}
    & Requirements   & Boundary Condition
     \\ \hline \vspace{-0.3cm} \\Type-I &
  $\left\{\begin{array}{l}
        |m|<|l_2+l_3|  \\
        l_4=0\\
        n(l_2+l_3)= 2l_1m \\
        l_2\neq l_3
   \end{array}\right.$  & 
   $\begin{array}{c}
   	\text{Both lower and upper boundary condition}\\ \text{(Eq.~(\ref{Boundary11}) and Eq.~(\ref{Boundary22})) }
   \end{array}$  \vspace{0.1cm} \\ \hline \vspace{-0.3cm} \\ Type-II
   & $\left\{\begin{array}{l}
        |m|<|l_2+l_3|  \\
        l_4=0\\
        n(l_2+l_3)\neq 2l_1m\\ l_2\neq l_3 
   \end{array}\right.$ or $\left\{\begin{array}{l}
        |m|<|l_2-l_3|  \\
        l_4\neq 0\\
        \det B =l_1l_4-l_2l_3=0\\
        (2l_2m-l_4n)(2l_3m-l_4n)=0
   \end{array}\right.$  &  $\begin{array}{l}
   \left\{\begin{array}{l}l_4=0\\|l_2|<|l_3|\end{array}\right.\text{ or }\left\{\begin{array}{l}l_4\neq 0\\2l_3m=l_4n\end{array}\right.,\ \text{Lower}~(\text{Eq.}~(\ref{Boundary11}))\\
        \left\{\begin{array}{l}l_4=0\\|l_2|>|l_3|\end{array}\right.\text{ or }\left\{\begin{array}{l}l_4\neq 0\\2l_2m=l_4n\end{array}\right.,\ \text{Upper}~(\text{Eq.}~(\ref{Boundary22})) 
   \end{array}$ \vspace{0.1cm} \\ \vspace{-0.4cm}
\end{tabular}
\end{ruledtabular}
\end{table*}

\section{Construction of iCS field theory and Toeplitz Braiding\label{sectionII}}

\subsection{iCS field theory with twisted $\mathbb{Z}_m$ topologically ordered layers\label{sectionIIB}}

The construction of an iCS field theory can be regarded as the process of stacking $\mathbb{Z}_m$ topological orders. We initiate our analysis from the $\mathbb{Z}_m$ twisted topological order, which is described by the $K$-matrix Chern-Simons theory $\mathcal{L} = \frac{1}{4\pi} \sum_{I,J} A_{IJ} a_\mu^I \partial_\nu a_\lambda^J \epsilon^{\mu\nu\lambda}$. Here, $A$ is specified as $A = \left(\begin{smallmatrix} n & m \\ m & 0 \end{smallmatrix}\right)$. Here, $a_\mu^I$ is a $U(1)$ gauge field and $n,m$ are quantized to integers due to the invariance under large gauge transformations. Given a stack of $\mathbb{Z}_m$ twisted topological order, we add Chern-Simons terms to couple the nearby layers, which corresponds to the off-diagonal blocks in the $K$ matrix. The stacking direction is dubbed the $z$-direction hereafter, and the resulting theory is described by the following Lagrangian:
\begin{equation}
    \mathcal{L} = \frac{1}{4\pi} \sum_{I,J} K_{IJ} a_\mu^I \partial_\nu a_\lambda^J \epsilon^{\mu\nu\lambda}, \label{lagrangianmcs}
\end{equation}
where 
\begin{equation}
    K = \left( \begin{matrix}
    A & B^T & & & & \\
    B & A & B^T & & & \\
    & B & A & B^T & & \\
    & & \ddots & \ddots & \ddots & \\
    & & & B & A & B^T \\
    & & & & B & A \\
    \end{matrix} \right), \label{ktridiagonal}
\end{equation}
with
	$A = \begin{pmatrix} n & m \\ m & 0 \end{pmatrix}$ and  $B = \begin{pmatrix} l_1 & l_2 \\ l_3 & l_4 \end{pmatrix}$, 
as shown in Eq.~(\ref{ktridiagonal}). 
Fig.~\ref{stackingZNtopo}(a) gives an illustration of the stacking process. Each layer is a $\ZZ_m$ twisted topological order described by $K$-matrix Chern-Simons theory corresponding to diagonal blocks $A$ in the Toeplitz $K$ matrix. The nearby layers are coupled by Chern-Simons terms, which corresponds to the upper-right and lower-left $2\times 2$ blocks $B$ in the $K$ matrix. For example, Layer 1 contributes intra-layer Chern-Simons terms $\frac{1}{4\pi}[m(a_\mu^1 \p_\nu a_\lambda^2+a_\mu^2 \p_\nu a_\lambda^1)+n (a_\mu^1\partial_\nu a_\lambda^1)]\epsilon^{\mu\nu\lambda}$. The inter-layer Chern-Simons terms between Layer 1 and Layer 2 are $\frac{1}{4\pi}[l_1(a_\mu^1 \p_\nu a_\lambda^3+a_\mu^3 \p_\nu a_\lambda^1)+l_2(a_\mu^3 \p_\nu a_\lambda^2+a_\mu^2 \p_\nu a_\lambda^3)+l_3(a_\mu^4 \p_\nu a_\lambda^1+a_\mu^1 \p_\nu a_\lambda^4)+l_4(a_\mu^4 \p_\nu a_\lambda^2+a_\mu^2 \p_\nu a_\lambda^4)]\epsilon^{\mu\nu\lambda}$. 
  If we stack $N$ layers of twisted $\mathbb{Z}_m$  topological order, the resulting theory is a multi-component Chern-Simons theory with a $2N$-dimensional $K$ matrix. 
  As the number of layers $N$ goes to infinity,\footnote{In this paper, taking thermodynamical limit refers to taking the limit $N\rightarrow\infty$ along $z$-direction. Each $xy$ layer is already in the thermodynamical limit.} the mCS theory becomes an iCS field theory as illustrated in Fig.~\ref{stackingZNtopo}(b). All the coefficients in the $K$ matrix are integers due to large gauge invariance. In this setting, the superscript $I$ of a gauge field $a_\mu^I$ carries the meaning of the $z$-coordinate. The $I$-th layer contains two independent gauge fields $a_\mu^{2I-1}$ and $a_\mu^{2I}$. We should be extremely careful about the general basis transformation $K \rightarrow W^\mT K W,\  W \in GL(2N,\mathbb{Z})$, requiring it to preserve the locality in the $z$-direction. The absence of upper-right and lower-left blocks indicates that we are discussing iCS field theories in $z$-OBC.

\begin{figure}
    \centering
\includegraphics[width=8cm]{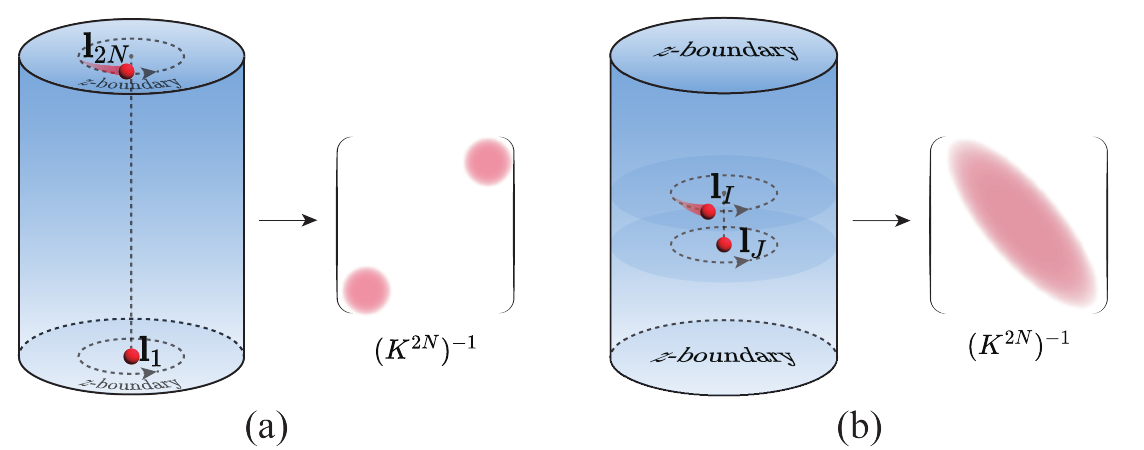}
    \caption{
    (a) is an illustrative diagram of braiding process between topological excitations
      residing at distinct $z$-boundaries. 
     In general, upper-right and lower-left elements of matrix $(K^{2N})^{-1}$ describe the braiding statistics.
     (b) is an illustrative diagram of braiding process between topological excitations $\mathbf l_I$ and $\mathbf l_J$ that are close in $z$-coordinates ($|I-J|\ll 2N$),
      which is described by the diagonal elements of matrix $(K^{2N})^{-1}$.}
    \label{braidingl1l2N}
\end{figure}

To discuss the braiding statistics between topological excitations, excitation terms need to be included in the Lagrangian Eq.~(\ref{lagrangianmcs}). In $K$-matrix Chern-Simons theories, topological excitations are labeled by integer vectors $\mathbf{p}=(p^1,p^2,\ldots,p^{2N})^\mT,\mathbf{q}=(q^1,q^2,\ldots,q^{2N})^\mT,\ldots\in\mathbb{Z}^{2N}$, and the excitation terms corresponding to $\mathbf{p}$ and $\mathbf{q}$ are given by $p^Ia_\mu^I j^\mu_{\mathbf{p}}$ and $q^Ia_\mu^I j^\mu_{\mathbf{q}}$, respectively, where $j^\mu_{\mathbf{p}}$ and $j^\mu_{\mathbf{q}}$ are the currents of topological excitations. After integrating out the gauge fields, the mutual braiding phase angle between topological excitations $\mathbf{p}$ and $\mathbf{q}$ is given by 
\begin{gather}
	\Theta_{\mathbf p,\mathbf q}=2\pi p^I (K^{-1})_{IJ}q^J \mod 2\pi. 
\end{gather}
This paper imposes a restriction on the mutual braiding phase angle, limiting it to the interval $[-\pi,\pi)$ unless otherwise specified. In the following text, we use $\mathbf{l}_I$, $I\in 1,\ldots, 2N$ to label the topological excitation carrying unit gauge charge, coupled to gauge field $a_\mu^I$. For example, $\mathbf{l}_1=(1,0,\ldots,0)^\mT$ labels an excitation coupled to the $a_\mu^1$ gauge field, residing at a $z$-boundary, while $\mathbf{l}_{2N}=(0,\ldots,0,1)^\mT$ labels an excitation residing at another $z$-boundary, coupled to gauge field $a_\mu^{2N}$. The mutual braiding phase angle between $\mathbf{l}_1$ and $\mathbf{l}_{2N}$ is $\Theta_{\mathbf l_1,\mathbf l_{2N}}=2\pi(K^{-1})_{1,2N}\mod 2\pi$. This example also tells us that the lower-left and upper-right elements of $K^{-1}$ describe the braiding phase angle between excitations residing at distinct $z$-boundaries, as illustrated in Fig.~\ref{braidingl1l2N}(a). On the contrary, the diagonal elements of $K^{-1}$ characterize the braiding statistics of excitations, where the distance between them is finite along the $z$-direction, as illustrated in Fig.~\ref{braidingl1l2N}(b).

\subsection{Boundary zero modes and Toeplitz braiding: A concrete example\label{zeromodetoeplitzbraiding}}
  Zero modes of $K$ matrices have the following properties: for a finite size $K$ matrix, the zero modes have finite values instead of being zero. As the size of the system, i.e., the size of the $K$ matrix, increases, the eigenvalues decrease to zero exponentially. Therefore, for a finite size $K$ matrix, the inverse $K^{-1}$ is always well-defined. In the following discussion, we mainly focus on the asymptotic behavior of $K^{-1}$ as the number of layers $N \rightarrow \infty$. Thus, practically, the inverse of $K$ is calculated before taking the limit of the system size\footnote{In this paper, system size refers to the dimension of a finite-size block-tridiagonal Toeplitz $K$ matrix.} $2N \rightarrow \infty$.

 To gain insights, we focus on a simple example, where the Lagrangian is given by Eq.~(\ref{lagrangianmcs}), and the $K$ matrix (denoted as $K_1$-matrix hereafter) is given by Eq.~(\ref{ktridiagonal}). The $A$ and $B$ blocks read:
\begin{gather}
    A=\begin{pmatrix}
        0& m\\ m & 0
    \end{pmatrix},\quad B=\begin{pmatrix}
        0&l\\ 0& 0
    \end{pmatrix},\label{icsssh}
\end{gather}
where $m,l\in\ZZ$. Denote the $K$ matrix of size $2N$ with block $A$, $B$ given by Eq.~(\ref{icsssh}) as $K_1^{2N}$, which is, in its mathematical form,  equivalent to  the Hamiltonian of Su-Schrieffer-Heeger model. 
The inverse $(K_1^{2N})^{-1}$ reads~\cite{movassagh2017green}
\begin{gather}
   (K_1^{2N})^{-1}=\notag \\ \resizebox{.9\hsize}{!}{$\begin{pmatrix}
        0&\frac{1}{m} & 0 & -\frac{l}{m^2} & 0 & \cdots & 0 & \frac{1}{m}\left(-\frac{l}{m}\right)^{N-1}\\
        \frac{1}{m} & 0 &0 & 0 & 0 &\cdots & 0 & 0\\
        0 & 0 & 0& \frac{1}{m} & 0 &  \cdots & 0 & \frac{1}{m}\left(-\frac{l}{m}\right)^{N-2}\\
        -\frac{l}{m^2} & 0 & \frac{1}{m} & 0 & 0 &  \cdots & 0  & 0\\
        \vdots & \vdots & \vdots & \vdots & \vdots & \ddots & \vdots & \vdots \\
        0 & 0 & 0 & 0 & 0 & \cdots & 0 & \frac{1}{m}\\
        \frac{1}{m}\left(-\frac{l}{m}\right)^{N-1} & 0 &  \frac{1}{m}\left(-\frac{l}{m}\right)^{N-2} & 0 & \frac{1}{m}\left(-\frac{l}{m}\right)^{N-3} & \cdots & \frac{1}{m} & 0
    \end{pmatrix}$},\label{ksshinverse}
\end{gather}
which captures the braiding statistics between topological excitations. Following the notation in Sec.~\ref{sectionIIB}, the topological excitation carrying unit gauge charge, coupled to gauge field $a_\mu^I$, is labelled by
\begin{equation}
    \bs l_I=\begin{pmatrix}
        \cdots &0&0&1&0&0&\cdots
    \end{pmatrix}^\mT.\label{unitcharge}
\end{equation}
The subscript $I$ refers to the location of the nonzero entry. 

\begin{figure}
    \centering
    \includegraphics[width=7cm]{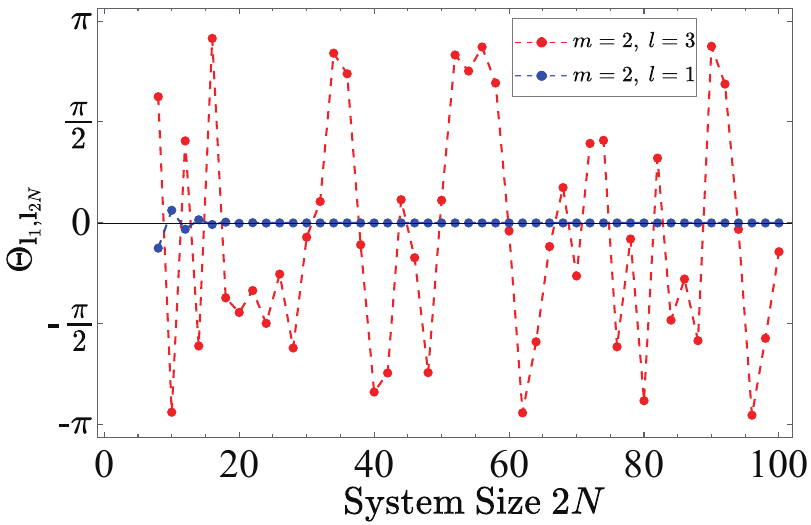}
    \caption{The mutual braiding phase angle $\Theta_{\mathbf l_1,\mathbf l_{2N}}=2 \pi \mathbf{l}_1^\mT (K_1^{2N})^{-1} \mathbf{l}_{2N}\mod 2\pi$, where we shift the phase angle in $(\pi,2\pi)$ to $(-\pi,0)$.  If $|m|>|l|$, the mutual braiding phase angle $\Theta_{\mathbf l_1,\mathbf l_{2N}}$ is suppressed as the system size $2N$ increases, where we take $m=2,l=1$ as an example.  In contrast, if $|m|<|l|$, the mutual braiding phase angle $\Theta_{\mathbf l_1,\mathbf l_{2N}}$ exhibits oscillation as the system size $2N$ increases. Here we take $m=2$, $l=3$ as an example.}
    \label{toep1}
\end{figure}

In the case $|m|>|l|$, the mutual braiding phase angle between $\mathbf l_I$ and $\mathbf l_J$ is $2\pi [(K_1^{2N})^{-1}]_{IJ}$, decaying exponentially with respect to $|I-J|$ for $I$, $J$ satisfying $I>J,\  I=2i ,\  J=2j-1,\ \text{or}\  J>I,\  J=2j,\  I=2i-1,\ i,j\in \ZZ^+$, which means that if two excitations reside at distinct $z$-boundaries,  they cannot  be mutually detected by braiding processes. For instance, the mutual braiding phase angle between two excitations $\bs l_1=\begin{pmatrix}
    1&0&\cdots&0
\end{pmatrix}^\mT$ and $\bs l_{2N}=\begin{pmatrix}
    0&\cdots&0&1
\end{pmatrix}^\mT$ is $\frac{2\pi}{m}\lb\frac{-l}{m}\rb^{N-1}\mod 2\pi$. Since $|l/m|<1$, the phase angle is suppressed to zero in thermodynamical limit $2N\rightarrow \infty$ as illustrated in Fig.~\ref{toep1}, where we take $m=2,l=1$ as an example.

If $|l|>|m|$,
the matrix element $ [(K_1^{2N})^{-1}]_{IJ}$
increases exponentially with $|I-J|$ for $I$, $J$ satisfying $I>J,\  I=2i ,\  J=2j-1,\ \text{or}\  J>I,\  J=2j,\  I=2i-1,\ i,j\in \ZZ^+$. The mutual braiding phase angle $\Theta_{\mathbf l_1,\mathbf l_{2N}}=\frac{2\pi}{m}\lb\frac{-l}{m}\rb^{N-1} \mod 2\pi$ between $\mathbf l_1$ and $\mathbf l_{2N}$ is not exponentially suppressed for $N$ being sufficiently large if $l$ is not divisible by $m~\mathrm{rad}(m)$\footnote{The radical $\mathrm{rad}(m)$ of an integer $m$ is the product of the distinct prime numbers dividing $m$.}, which indicates that topological excitations $\mathbf l_1$ and $\mathbf l_{2N}$ can be remotely detected by each other in the braiding process.
As the system size increases, the braiding phase angle between two excitations located at distinct $z$-boundaries, $\mathbf l_1$ and $\mathbf l_{2N}$, exhibits oscillation, as shown in Fig.~\ref{toep1}. 

Furthermore, we will demonstrate that the presence of  \textit{Toeplitz braiding} requires  boundary zero modes. As is known, if $|l|>|m|$, the $K_1^{2N}$ matrix has boundary zero modes $\mathbf{v}_1$ and $\mathbf{v}_2$,
\begin{subequations}
	\begin{gather}
    \mathbf v_1=\frac{1}{\sqrt{2}} \mathbf u_1 +\frac{1}{\sqrt{2}}\mathbf u_2,\label{rococo1}\\
    \mathbf v_2=-\frac{1}{\sqrt{2}} \mathbf u_1 +\frac{1}{\sqrt{2}}\mathbf u_2,\label{rococo2}
\end{gather}
\end{subequations}
where
\begin{subequations}
	\begin{gather}
\begin{split}
    \mathbf u_1=&\sqrt{\frac{1-\left(\frac{m}{l}\right)^2}{1-\left(\frac{m}{l}\right)^{2N}}}\\ &\left(1 , 0 , \left(-\frac{m}{l}\right) , 0 , \left(-\frac{m}{l}\right)^2 , \cdots , \left(-\frac{m}{l}\right)^{N-1} , 0\right)^T,
\end{split}\label{monalisa1}\\
    \begin{split}
    \mathbf u_2=&\sqrt{\frac{1-\left(\frac{m}{l}\right)^2}{1-\left(\frac{m}{l}\right)^{2N}}}\\ &\left(0 , \left(-\frac{m}{l}\right)^{N-1} , 0 , \left(-\frac{m}{l}\right)^{N-2} , \cdots , \left(-\frac{m}{l}\right) , 0 , 1\right)^T.
\end{split}\label{monalisa2}
\end{gather}
\end{subequations}
Only when the system size $2N\rightarrow \infty$, the eigenvalues of boundary zero modes are strictly equal to zero. The eigenvalues of $\mathbf{v}_{1}$ and $\mathbf{v}_{2}$ are denoted as $\lambda_{1}$ and $\lambda_{2}$, respectively. 
\begin{gather}
    \lambda_1=m\frac{1-\left(\frac{m}{l}\right)^2}{1-\left(\frac{m}{l}\right)^{2N}}\left(-\frac{m}{l}\right)^{N-1},\\  \lambda_2= -m\frac{1-\left(\frac{m}{l}\right)^2}{1-\left(\frac{m}{l}\right)^{2N}}\left(-\frac{m}{l}\right)^{N-1}.
\end{gather}
The inverse of $K_1^{2N}$ can be written as
\begin{equation}
\begin{split}
        (K_1^{2N})^{-1}=&\frac{1}{\lambda_1}\mathbf{v}_1\mathbf{v}_1^\dag+\frac{1}{\lambda_{2}}\mathbf{v}_2\mathbf{v}_2^\dag+\frac{1}{\lambda_3}\mathbf{v}_3\mathbf{v}_3^\dag+\cdots\\&+\frac{1}{\lambda_{2N}}\mathbf{v}_{2N}\mathbf{v}_{2N}^\dag,
\end{split}
\end{equation}
$\mathbf{v}_3,\ldots,\mathbf{v}_{2N}$ are the other eigenvectors of $K_1^{2N}$, and $\lambda_3,\ldots,\lambda_{2N}$ are the corresponding eigenvalues.
As far as $(K_1^{2N})^{-1}$ is concerned, $\frac{ 1}{\lambda_1}$ and $\frac{ 1}{\lambda_2}$ are eigenvalues of $(K_1^{2N})^{-1}$ with largest absolute values. In the thermodynamical limit $2N\rightarrow \infty$, $\frac{ 1}{\lambda_1}$ and $\frac{ 1}{\lambda_2}$ diverge. We wonder what information is captured in
\begin{equation}
    M_1^{2N}:=\frac{1}{\lambda_1}\mathbf{v}_1\mathbf{v}_1^\dag+\frac{1}{\lambda_{2}}\mathbf{v}_2\mathbf{v}_2^\dag. \label{m12n}
\end{equation}
It should be noted that the rank of $M_1^{2N}$ is 2, thus it is not invertible. 
Having $\mathbf v_1$ and $\mathbf v_2$, we are able to write down the matrix $M_1$
\begin{gather}
    M_1^{2N}=\frac{1}{\lambda_1} \mathbf{v}_1 \mathbf{v}_1^\dag+\frac{1}{\lambda_2} \mathbf{v}_2 \mathbf{v}_2^\dag= \notag \\
    \resizebox{\hsize}{!}{$\begin{pmatrix}
        0&\frac{1}{m} & 0 & -\frac{l}{m^2} & 0 & \cdots & \frac{1}{m}\left(-\frac{l}{m}\right)^{N-2} & 0 & \frac{1}{m}\left(-\frac{l}{m}\right)^{N-1}\\
        \frac{1}{m} & 0 &-\frac{1}{l} & 0 & \frac{m}{l^2} &\cdots & 0 & \frac{1}{m}\left(-\frac{m}{l}\right)^{N-1} & 0\\
        0 & -\frac{1}{l} & 0& \frac{1}{m} & 0 &  \cdots & \frac{1}{m}\left(-\frac{l}{m}\right)^{N-3} & 0 & \frac{1}{m}\left(-\frac{l}{m}\right)^{N-2}\\
        \vdots & \vdots & \vdots & \vdots & \vdots & \ddots & \vdots & \vdots & \vdots \\
        \frac{1}{m}\left(-\frac{l}{m}\right)^{N-2} & 0 & \frac{1}{m}\left(-\frac{l}{m}\right)^{N-3} & 0 & \frac{1}{m}\left(-\frac{l}{m}\right)^{N-4} &  \cdots & 0 & -\frac{1}{l}  & 0\\
        0 & \frac{1}{m}\left(-\frac{m}{l}\right)^{N-1} & 0 & \frac{1}{m}\left(-\frac{m}{l}\right)^{N-2} & 0 & \cdots & -\frac{1}{l} & 0 & \frac{1}{m}\\
        \frac{1}{m}\left(-\frac{l}{m}\right)^{N-1} & 0 &  \frac{1}{m}\left(-\frac{l}{m}\right)^{N-2} & 0 & \frac{1}{m}\left(-\frac{l}{m}\right)^{N-3} & \cdots & 0 & \frac{1}{m} & 0
    \end{pmatrix}.$}\label{colossus}
\end{gather}
The upper-right and lower-left elements $(M_1^{2N})_{IJ}=\frac{1}{m}\left(-\frac{m}{l}\right)^{(|I-J|+1)/2},\ I= 2i,\ J= 2j-1,\ i,j\in\ZZ^+$ are exponentially suppressed. Therefore, $M_1^{2N}$ gives good approximation to the upper-right and lower-left elements of $({K_1^{2N}})^{-1}$. In other words, $M_1^{2N}$ captures the long-range braiding statistics in $z$-direction.

\begin{figure}
    \centering
    \includegraphics[width=8cm]{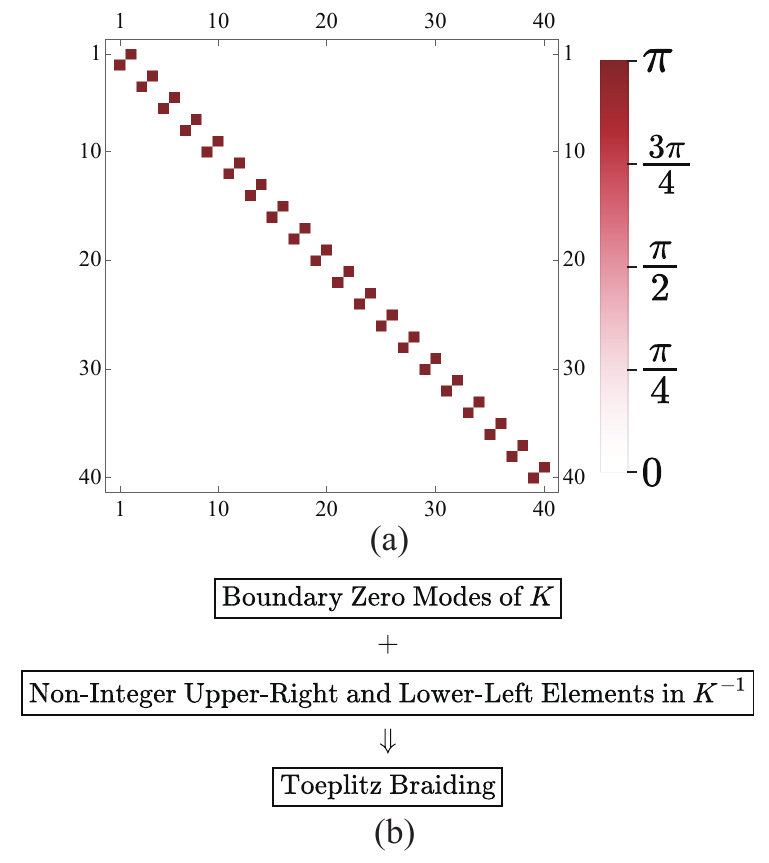}
    \caption{ (a) is the matrix plot of mutual braiding phase angle $\Theta_{\mathbf l_I,\mathbf l_J}=2\pi (K_1^{2N})^{-1}_{IJ}$ with $m=2$, $l=4$, $2N=40$, where the mutual braiding phase angle is limited to $(-\pi,\pi]$. Nontrivial braiding statistics is completely absent except for the topological excitations residing at the same layer. (b) illustrates the conditions for Toeplitz braiding.}
    \label{conditionToeplitz}
\end{figure}


It is noteworthy that in some cases, the nonzero upper-right and lower-left elements may not result in nontrivial Toeplitz braiding. For example, if $l$ is divisible by $m~\mathrm{rad}(m)$, the upper-right and lower-left elements in Eq.~(\ref{ksshinverse}) are integers, implying that excitations residing at distinct $z$-boundaries cannot interact by braiding. \textit{Indeed, the presence of boundary zero modes is a necessary but not sufficient condition for the appearance of Toeplitz braiding.} We have demonstrated that the appearance of boundary zero modes in $K$ matrices will result in non-zero elements in the upper-right and lower-left regions of $K^{-1}$. However, the elements in these regions are not guaranteed to be non-integers. The $K_1$-type iCS field theory with $m=2$, $l=4$ is a typical example. The $K$ matrix has boundary zero modes, but nontrivial braiding statistics is completely absent except for the topological excitations residing at the same layer, as shown in Fig.~\ref{conditionToeplitz}(a). The presence of non-integers in the upper-right and lower-left regions of $K^{-1}$ is necessary for the occurrence of Toeplitz braiding. Therefore, we propose the following condition for the existence of Toeplitz braiding: \textit{If boundary zero modes of $K$ along with non-integer upper-right and lower-left regions of $K^{-1}$ exist, then Toeplitz braiding occurs}, which is also illustrated in Fig.~\ref{conditionToeplitz}(b).

\section{Sorting $K$ matrices with boundary zero modes\label{sectionIII}}
\subsection{General approach to sorting $K$ matrices with boundary zero modes\label{sectionIIIA}}

Our investigation into a concrete example of iCS field theories with Toeplitz braiding in Sec.~\ref{sectionII} raises an intriguing question: how to identify and classify iCS field theories with Toeplitz braiding? As illustrated in Fig.~\ref{conditionToeplitz}(b), the appearance of Toeplitz braiding demands both boundary zero modes of $K$ and non-integer upper-right and lower-left elements in $K^{-1}$. In this section, we present a systematic framework for determining the presence of boundary zero modes in a given $K$ matrix. As shown in Table \ref{icstoeplitz}, we identify two different types of $K$ matrices with boundary zero modes as illustrative examples. The Type-I $K$ matrix possesses two linearly independent boundary zero modes, whereas the Type-II $K$ matrix exhibit only one boundary zero mode. The corresponding iCS field theories, are dubbed as Type-I and Type-II iCS field theories, respectively.  The aforementioned example, an iCS field theory with an Su-Schrieffer-Heeger (SSH)-type $ K$-matrix, represents a typical Type-I iCS field theory.

Denote the entries of an eigenvector $\mathbf{u}$ of $K$ satisfying the eigenequation $K\mathbf{u}=\lambda \mathbf{u}$ as
\begin{gather}
    \mathbf{u}=\begin{pmatrix}
    u_1^a & u_1^b & u_2^a & u_2^b & \cdots & u_N^a & u_N^b
\end{pmatrix}^T.
\label{ansatzpsi1sb}
\end{gather}
Since we are discussing boundary zero modes, the $\lambda\rightarrow 0$ limit is taken at first. In terms of  the Toeplitz $K$ matrices in the form of Eq.~(\ref{ktridiagonal}), the entries of a trial solution $\mathbf u$ to boundary zero mode are required to satisfy the recurrence relation
\begin{gather}
    B \begin{pmatrix}
        u_{j-1}^a \\ u_{j-1}^b
    \end{pmatrix}+A\begin{pmatrix}
        u_j^a \\ u_j^b
    \end{pmatrix}+B^\mT \begin{pmatrix}
        u_{j+1}^a \\ u_{j+1}^b
    \end{pmatrix}=0\label{Bulk12}
\end{gather}
for $j=2,\ldots, N-1$. Moreover, $\mathbf u$ is required to satisfy the ``boundary conditions"
\begin{subequations}
    \begin{equation}
        A\begin{pmatrix}
        u_1^{a} \\ u_1^{b}
    \end{pmatrix}+B^\mT \begin{pmatrix}
        u_{2}^{a} \\ u_{2}^{b}
    \end{pmatrix}=0,\label{Boundary11}
    \end{equation}\vspace{-0.5cm}
    \begin{equation}
         B \begin{pmatrix}
        u_{N-1}^{a} \\ u_{N-1}^{b}
    \end{pmatrix}+A\begin{pmatrix}
        u_N^{a} \\ u_N^{b}
    \end{pmatrix}=0.\label{Boundary22}
    \end{equation}
\end{subequations}
Eq.~(\ref{Boundary11}) and Eq.~(\ref{Boundary22}) are also dubbed   ``lower boundary condition" and ``upper boundary condition", respectively.
To solve the recurrence relation Eq.~(\ref{Bulk12}) as well as the boundary conditions Eq.~(\ref{Boundary11}) and Eq.~(\ref{Boundary22}), we take the following ansatz
\begin{gather}
    \mathbf u=\sum_j c_j \mathbf w^{j},\label{superpositionbasis}
\end{gather}
where $\{c_j\}$ are coefficients yet to be determined and $\{\mathbf w^j\}$ are dubbed as ``boundary basis". We assume that the a boundary base $\mathbf w^j$ has the form
\begin{gather}
    \resizebox{\hsize}{!}{$\mathbf w^j=\begin{pmatrix}
         \beta_jw^{j,a} & \beta_jw^{j,b} & \beta_j^2 w^{j,a} & \beta_j^2 w^{j,b} & \cdots & \beta_j^{N} w^{j,a} & \beta_j^{N} w^{j,b}
    \end{pmatrix}^T$}. \label{phij123}
\end{gather}
 The recurrence relation Eq.~(\ref{Bulk12}) gives the equation for $\begin{pmatrix}
    w^{j,a} & w^{j,b}
\end{pmatrix}^\mT$,
\begin{gather}
    B \begin{pmatrix}
        w^{j,a} \\ w^{j,b}
    \end{pmatrix}+ \beta_j A\begin{pmatrix}
        w^{j,a} \\ w^{j,b}
    \end{pmatrix}+\beta_j^2 B^\mT \begin{pmatrix}
        w^{j,a} \\ w^{j,b}
    \end{pmatrix}=0.\label{Bulk1phi}
\end{gather}
Nontrivial solution to $\begin{pmatrix}
    w^{j,a} & w^{j,b}
\end{pmatrix}^\mT$ renders the equation for $\beta_j$
\begin{gather}
    \det(B+A\beta_j+B^\mT\beta_j^2)=0.
\end{gather}
Therefore, $\{\beta_j\}$ are the solutions to the equation
\begin{gather}
\alpha_1+\alpha_2\beta+\alpha_3\beta^2+\alpha_2\beta^3+\alpha_1\beta^4=0 , \label{equationforbeta11ps} \\ \left\{
\begin{array}{l}
     \alpha_1=l_1l_4-l_2l_3, \\ \alpha_2=-(m(l_2+l_3)-nl_4),\\
     \alpha_3=-(l_2^2+l_3^2-2l_1l_4+m^2).
\end{array}\right.
\end{gather}
$\beta_j$ together with the normalization condition $(\mathbf w^{j})^\dagger \mathbf w^{j}=1$ settles down the entries of $\mathbf w^j$.
Furthermore, we require that $\mathbf{u}$ localizes at one $z$-boundary, i.e., all the  $\{\beta_j\}$ in one trial solution $\mathbf{u}$ satisfies either $|\beta_j|>1$ or $|\beta_j|<1$.
 If a trial solution $\mathbf{u}$ is a linear combination of boundary bases $\{\mathbf w^j\}$ with $|\beta_j|<1$, the boundary condition Eq.~(\ref{Boundary22}) is automatically satisfied in the thermodynamical limit $2N\rightarrow\infty$. Hence our task is to verify whether a linear combination of boundary bases (given by Eq.~(\ref{superpositionbasis})) exists, matching the lower boundary condition Eq.~(\ref{Boundary11}). Likewise, for boundary bases $\{\mathbf w^j\}$ with $|\beta_j|>1$, our task is to verify whether a linear combination of boundary bases matching the upper boundary condition Eq.~(\ref{Boundary22}) exists. If $\mathbf u$ satisfies the lower boundary condition Eq.~(\ref{Boundary11}), it is referred to as lower boundary zero mode. If $\mathbf u$ satisfies the upper boundary condition Eq.~(\ref{Boundary22}), it is referred to as upper boundary zero mode.
Following this framework, we can systematically determine whether a given $K$ matrix possesses boundary zero modes. 

In the preceding discussion, we made the assumption that the $K$ matrix is infinite-dimensional. However, in the study of  \textit{Toeplitz braiding}, we calculate the inverse of a $K$ matrix before taking the thermodynamic limit $2N\rightarrow\infty$. Therefore, it becomes essential to incorporate  the discussion on the boundary zero modes and the corresponding exponentially-suppressed eigenvalues of a \textit{finite-size} $K$ matrix. If an infinite-dimensional $K$ matrix possesses two boundary zero modes, $\mathbf u^1$ and $\mathbf u^2$, satisfying distinct boundary conditions Eq.~(\ref{Boundary11}) and Eq.~(\ref{Boundary22}), respectively, the actual boundary zero modes for a \textit{finite-size} $K$ matrix are linear combinations of $\mathbf u^1$ and $\mathbf u^2$, which can be obtained by diagonalizing the matrix
\begin{gather}
    \begin{pmatrix}
      (\mathbf u^1)^\dag K \mathbf u^1 & (\mathbf u^1)^\dag K \mathbf u^2 \\
      (\mathbf u^2)^\dag K \mathbf u^1 & (\mathbf u^2)^\dag K \mathbf u^2
    \end{pmatrix}.\label{twozeromodeactual}
\end{gather}
For instance, for a finite-size $K_1$-type matrix, the boundary zero modes $\mathbf{u}^1$ and $\mathbf{u}^2$ meeting distinct boundary conditions are presented in Eq.~(\ref{monalisa1}) and Eq.~(\ref{monalisa2}). The actual boundary zero modes $\mathbf v_1$, $\mathbf v_2$ (Eq.~(\ref{rococo1}) and  Eq.~(\ref{rococo2})) and the corresponding exponentially suppressed eigenvalues of $K_1^{2N}$, are obtained by diagonalizing
\begin{gather}
\begin{split}
	&\begin{pmatrix}
      (\mathbf u^1)^\dag K_1^{2N} \mathbf u^1 & (\mathbf u^1)^\dag K_1^{2N} \mathbf u^2 \\
      (\mathbf u^2)^\dag K_1^{2N} \mathbf u^1 & (\mathbf u^2)^\dag K_1^{2N} \mathbf u^2
    \end{pmatrix}\\
    =&\frac{1-\left(\frac{m}{l}\right)^2}{1-\left(\frac{m}{l}\right)^{2N}}\begin{pmatrix}
      0 & m \left(-\frac{m}{l}\right)^{N-1} \\
      m \left(-\frac{m}{l}\right)^{N-1} & 0
    \end{pmatrix}.
\end{split}
\end{gather}
Having $\mathbf v_1$, $\mathbf v_2$ and the exponentially suppressed eigenvalues $\lambda_1$, $\lambda_2$, we are able to construct the matrix $M_1^{2N}$ (Eq.~(\ref{m12n})), which captures the braiding statistics between topological excitations residing at distinct $z$-boundaries.

If an infinite-dimensional $K$ matrix possesses only one boundary zero mode $\mathbf u$ meeting either Eq.~(\ref{Boundary11}) or Eq.~(\ref{Boundary22}), it is in fact not an actual boundary zero mode for a \textit{finite-size} $K$ matrix due to the violation to the other boundary condition. The violation will provide a minor correction to the entries of $\mathbf u$, contributing a small but nonzero eigenvalue $\lambda$, which enables the calculation of the inverse of $K$. In Sec.~\ref{sectionIV}, we will demonstrate that $M=\frac{1}{\lambda}\mathbf u \mathbf u^\dag$ also captures the braiding statistics of topological excitations residing at distinct $z$-boundaries, in analogy to the cases of two boundary zero modes. 

In the subsequent subsections, we provide a detailed discussion on sorting the Type-I and  Type-II Toeplitz $K$ matrices. 
For a $K$ matrix as shown in Eq.~(\ref{ktridiagonal}), we follow a two-step procedure. First, we solve for all the boundary bases, then we verify whether a linear superposition of these boundary bases exists, as expressed in Eq.~(\ref{superpositionbasis}), satisfying either the upper boundary condition Eq.~(\ref{Boundary11}) or lower boundary condition Eq.~(\ref{Boundary22}). Therefore,
for a given (twisted) topological order described by $A=\begin{pmatrix} n & m\\ m& 0 \end{pmatrix}$, we can identify the specific Chern-Simons coupling associated with the block $B$ in the $K$ matrix that enables Toeplitz braiding.
We organize our analysis based on whether the determinant of matrix $B$ is equal to zero or not. $\det B$ plays a crucial role in determining the order of Eq.~(\ref{equationforbeta11ps}), which affects the \# of boundary bases.

\subsection{Discussion on block $B$}
If $\det B=\alpha_1=0$, Eq.~(\ref{equationforbeta11ps}) has a solution $\beta_0=0$, which is discarded. The other two solutions $\beta_1$ and $\beta_2$, are determined by the equation
\begin{gather}
    \alpha_2+\alpha_3 \beta_{1,2}+\alpha_2 \beta_{1,2}^2=0,
\end{gather}
from which we know that $\beta_1\beta_2=1$. For convenience, we assume $|\beta_1|<1$, $|\beta_2|>1$. The bulk recurrence relation  Eq.~(\ref{Bulk1phi}) together with normalization condition gives the solution to $\begin{pmatrix}
    w^{j,a} & w^{j,b}
\end{pmatrix}^T$, $j=1,2$. If $\mathbf{u}^1=\mathbf w^{1}$ is a legitimate boundary zero mode, it satisfies the boundary condition Eq.~(\ref{Boundary11}). 
  Detailed calculation in Appendix \ref{section1} shows that there exists a boundary zero mode satisfying the lower boundary condition Eq.~(\ref{Boundary11}) if the condition (\ref{godel1}) is satisfied. 
\begin{widetext}
    \begin{gather}
    \mathbb A_1: \left\{\begin{array}{l}
    m\neq 0,\quad l_4\neq 0,\quad l_2\neq l_3,\\
    l_1l_4=l_2l_3,\quad
    2l_3m=l_4n,\\
    |m|<|l_3-l_2|,
    \end{array}\right. 
    \text{ or } \mathbb A_2: \left\{\begin{array}{l}
    m\neq 0,\quad l_3\neq 0,\\
    l_2=l_4=0,\\
    |m|<|l_3|,
    \end{array}\right.\text{ or } \mathbb A_3:\left\{\begin{array}{l}
    m\neq 0,\quad l_2\neq 0,\\
    nl_2=2l_1m,\\
    l_3=l_4=0,\\
    |m|<|l_2|.
    \end{array}\right. \label{godel1} 
\end{gather}
\end{widetext}
Likewise, there exists a boundary zero mode $\mathbf{u}^2=\mathbf w^2$ satisfying the upper boundary condition Eq.~(\ref{Boundary22}) if the boundary condition (\ref{godel2}) is satisfied. 
\begin{widetext}
    \begin{gather}
 \mathbb B_1:
        \left\{\begin{array}{l}
    m\neq 0,\quad l_4\neq 0,\quad l_2\neq l_3,\\
    l_1l_4=l_2l_3,\quad
    2l_2m=l_4n,\\
    |m|<|l_3-l_2|,\\
    \end{array}\right.  \text{ or } \mathbb B_2: \left\{\begin{array}{l}
    m\neq 0,\quad l_2\neq 0,\\
    l_3=l_4=0,\\
    |m|<|l_2|,
    \end{array}\right.\text{ or } \mathbb B_3: \left\{\begin{array}{l}
    m\neq 0,\quad l_3\neq 0,\\
    nl_3=2l_1m,\\
    l_2=l_4=0,\\
    |m|<|l_3|.
    \end{array}\right.\label{godel2}
\end{gather}
\end{widetext}
Here Eq.~(\ref{godel1}) represents the union $\mathbb A_1\cup \mathbb A_2\cup \mathbb A_3$ of the sets $\mathbb A_{1,2,3}$. Likewise, Eq.~(\ref{godel2}) represents the union $\mathbb B_1 \cup \mathbb B_2 \cup \mathbb B_3$. Intersecting the above conditions Eq.~(\ref{godel1}) and Eq.~(\ref{godel2}), i.e. the intersection $(\mathbb A_1\cup \mathbb A_2\cup \mathbb A_3)\cap(\mathbb B_1\cup \mathbb B_2\cup \mathbb B_3)$, we should have 9 sets of conditions $\mathbb A_{1,2,3}\cap \mathbb B_{1,2,3}$ to ensure that the $K$ matrix has two boundary zero modes. With further analysis, only $\mathbb A_3 \cap \mathbb B_2$ and $\mathbb A_2\cap \mathbb B_3$ are not empty set, where the remaining sets are empty set. Therefore, we conclude that if $\det B=0$, and the elements of the $K$ matrix satisfies 
\begin{gather}
	\left\{\begin{array}{l}
    m\neq 0,\quad l_2\neq 0,\\
    l_3=l_4=0,\\
    nl_2=2l_1m,\\
    |m|<|l_2|,
    \end{array}\right.\text{ or }\left\{\begin{array}{l}
    m\neq 0,\quad l_3\neq 0,\\
    l_2=l_4=0,\\
    nl_3=2l_1m,\\
    |m|<|l_3|,
    \end{array}\right.\label{feizhu}
\end{gather}
the $K$ matrix exhibits two boundary zero modes. Otherwise, if $\det B=0$, and the elements of Eq.~(\ref{ktridiagonal}) satisfies only one of the two conditions Eq.~(\ref{godel1}) and Eq.~(\ref{godel2}), the $K$ matrix possesses only one boundary zero mode. Furthermore, the conditions for only one boundary zero mode under $\det B=0$ are summarized as follows:
	\begin{gather}
    \left\{\begin{array}{l}
    m\neq 0,\quad l_4\neq 0,\\
    l_1l_4=l_2l_3,\\
    2l_3m=l_4n\ \text{ or }\ 2l_2m=l_4n,\\
    |m|<|l_3-l_2|,
    \end{array}\right.\text{ or }\left\{\begin{array}{l}
    m\neq 0,\quad
    l_3\neq 0,\\
    l_4=l_2= 0,\\
    nl_3 \neq 2l_1 m\\
         |m|<|l_3|,
    \end{array}\right.\notag\\
    \text{ or }\left\{\begin{array}{l}
    m\neq 0,\quad
    l_2\neq 0,\\
    l_4=l_3= 0,\\
    nl_2 \neq 2l_1 m\\
         |m|<|l_2|.
    \end{array}\right. \label{lifesucks}
\end{gather}
Eq.~(\ref{lifesucks}) is obtained by the set operation $ (\mathbb A_1\cup \mathbb A_2\cup \mathbb A_3\cup\mathbb B_1\cup \mathbb B_2\cup \mathbb B_3 )\backslash [(\mathbb A_1\cup \mathbb A_2\cup \mathbb A_3)\cap(\mathbb B_1\cup \mathbb B_2\cup \mathbb B_3)]$.
More specifically, if $\left\{\begin{array}{l}l_4=0\\|l_2|<|l_3|\end{array}\right.\text{ or }\left\{\begin{array}{l}l_4\neq 0\\2l_3m=l_4n\end{array}\right.$, the boundary zero mode is a lower boundary zero mode, satisfying the boundary condition Eq.~(\ref{Boundary11}). If $\left\{\begin{array}{l}l_4=0\\|l_3|<|l_2|\end{array}\right.\text{ or }\left\{\begin{array}{l}l_4\neq 0\\2l_2m=l_4n\end{array}\right.$, the boundary zero mode is an upper boundary zero mode, satisfying the boundary condition Eq.~(\ref{Boundary22}). 

 If $\alpha_1=\det B\neq 0$, Eq.~(\ref{equationforbeta11ps}) is a fourth-degree equation. It has 4 solutions $\beta_1$, $\beta_2$, $\beta_3$, $\beta_4$, satisfying $\beta_1\beta_2\beta_3\beta_4=1$. Here the solutions $\beta_1$, $\beta_2$ satisfies $|\beta_1|<1$, $|\beta_2|<1$ while the other two solutions satisfies $|\beta_3|>1$, $|\beta_4|>1$.\footnote{The definitions of $\beta_1$, $\beta_2$, $\beta_3$, $\beta_4$ are different from those presented in Appendix \ref{section2}.} 
Therefore, a boundary zero mode satisfying boundary condition Eq.~(\ref{Boundary11}) is a linear combination of $\mathbf w^1$ and $\mathbf w^2$ given by Eq.~(\ref{phij123}). Likewise, a boundary zero mode satisfying boundary condition (\ref{Boundary22}) is a linear combination of $\mathbf w^3$ and $\mathbf w^4$ given by Eq.~(\ref{phij123}).
In general, analyzing the quartic equation Eq.~(\ref{equationforbeta11ps}) is very challenging. It often requires a case-by-case study to determine whether a given $K$ matrix possesses boundary zero modes. However, if $l_4=0$, Eq.~(\ref{equationforbeta11ps}) can be factorized into the product of two quadratic polynomials,
\begin{gather}
    (l_2+m\beta+l_3\beta^2)(l_3+m\beta+l_2\beta^2)=0,
\end{gather}
which enables a thorough discussion.
Detailed calculation in Appendix \ref{section2} shows that there exists a boundary zero mode satisfying the lower boundary condition Eq.~(\ref{Boundary11}) if the following conditions (\ref{macbook1}) are satisfied.
\begin{gather}
	\mathbb C_1:\ \left\{
\begin{array}{l}
l_4=0,\quad n(l_2+l_3)=2l_1 m,\\
m\neq 0,\quad  l_2l_3\neq 0, \\
|m|<|l_2+l_3|,\quad |l_3|<|l_2|,\\
\end{array}
\right.\quad \text{or} \notag \\ 
\mathbb C_2:\ \left\{\begin{array}{l}
	    l_4=0,\\
		m\neq 0,\quad  l_2l_3\neq 0,\\
		|m|<|l_2+l_3|,\quad |l_2|<|l_3|.
	\end{array}\right.\label{macbook1}
\end{gather}
Likewise, there exists one boundary zero mode satisfying the upper boundary condition Eq.~(\ref{Boundary22}) if the following conditions (\ref{macbook2}) are satisfied.
\begin{gather}
	\mathbb D_1:\ \left\{\begin{array}{l}
	    l_4=0,\\
		m\neq 0,\quad  l_2l_3\neq 0,\\
		|m|<|l_2+l_3|,\quad |l_3|<|l_2|,
	\end{array}\right. \quad \text{or} \notag \\
	\mathbb D_2:\ \left\{
\begin{array}{l}
l_4=0,\quad n(l_2+l_3)=2l_1 m,\\
m\neq 0,\quad  l_2l_3\neq 0, \\
|m|<|l_2+l_3|,\quad |l_2|<|l_3|.\\
\end{array}
\right. \label{macbook2}
\end{gather}
Here Eq.~(\ref{macbook1}) and Eq.~(\ref{macbook2}) represents the union $\mathbb C_1 \cup \mathbb C_2$ and $\mathbb D_1 \cup \mathbb D_2$, respectively.
Set operation $(\mathbb C_1 \cup \mathbb C_2)\cap(\mathbb D_1 \cup \mathbb D_2)$ yields the following condition that ensures the $K$ matrix (Eq.~(\ref{ktridiagonal})) possessing two boundary zero modes:
\begin{gather}
    \left\{\begin{array}{l}
    l_4=0,\\
        |m|<|l_2+l_3|,  \\
        n(l_2+l_3)= 2l_1m,\\ l_2\neq l_3,\quad l_2l_3 \neq 0.
   \end{array}\right. \label{kavakos2}
\end{gather}
Furthermore, set operation $(\mathbb C_1 \cup \mathbb C_2\cup\mathbb D_1 \cup \mathbb D_2)\backslash [(\mathbb C_1 \cup \mathbb C_2)\cap(\mathbb D_1 \cup \mathbb D_2)]$ yields the following condition for the appearance of exactly one boundary zero mode:
\begin{gather}
    \left\{\begin{array}{l}
    l_4=0,\\
        |m|<|l_2+l_3|,  \\
        n(l_2+l_3)\neq 2l_1m,\\ l_2\neq l_3,\quad l_2l_3 \neq 0.
   \end{array}\right. \label{kavakos1}
\end{gather}
More specifically, for the case $|l_2|<|l_3|$, the boundary zero mode satisfies the lower boundary condition Eq.~(\ref{Boundary11}); for the case $|l_2|>|l_3|$, the boundary zero mode satisfies the upper boundary condition Eq.~(\ref{Boundary22}).

\subsection{Condition for Type-I and Type-II iCS field theories}
Taking the union of (\ref{godel1}) and (\ref{macbook1}), we find that the $K$ matrix Eq.~(\ref{ktridiagonal}) has a boundary zero mode matching the boundary condition Eq.~(\ref{Boundary11}) if one of the following conditions is satisfied:
\begin{widetext}
\begin{gather}
	\mathbb M_1:~\left\{\begin{array}{l}
    m\neq 0,\quad l_4\neq 0,\\
    l_1l_4=l_2l_3,\quad
    2l_3m=l_4n,\\
    |m|<|l_3-l_2|,
    \end{array}\right. 
    \quad \text{or} \quad 
   \mathbb M_2: \left\{
\begin{array}{l}
m\neq 0, \\
l_4=0,\quad n(l_2+l_3)=2l_1 m,\\
|m|<|l_2+l_3|,\quad |l_3|<|l_2|,\\
\end{array}
\right.\quad \text{or} \quad \mathbb M_3:
\left\{\begin{array}{l}
		m\neq 0,\\
		l_4=0,\\
		|m|<|l_2+l_3|,\quad |l_2|<|l_3|.
	\end{array}\right.
\end{gather}
\end{widetext}
Similarly, taking the union of (\ref{godel2}) and (\ref{macbook2}), we find that the $K$ matrix Eq.~(\ref{ktridiagonal}) has a boundary zero mode matching the boundary condition Eq.~(\ref{Boundary22}) if one of the following conditions is satisfied:
\begin{widetext}
\begin{gather}
	\mathbb N_1:~\left\{\begin{array}{l}
    m\neq 0,\quad l_4\neq 0,\\
    l_1l_4=l_2l_3,\quad
    2l_2m=l_4n,\\
    |m|<|l_3-l_2|,
    \end{array}\right. 
    \quad \text{or} \quad \mathbb N_2:
    \left\{
\begin{array}{l}
m\neq 0, \\
l_4=0,\quad n(l_2+l_3)=2l_1 m,\\
|m|<|l_2+l_3|,\quad |l_2|<|l_3|,\\
\end{array}
\right.\quad \text{or} \quad \mathbb N_3:
\left\{\begin{array}{l}
		m\neq 0,\\
		l_4=0,\\
		|m|<|l_2+l_3|,\quad |l_3|<|l_2|.
	\end{array}\right.
\end{gather}
\end{widetext}
The set operation $(\mathbb M_1 \cup \mathbb M_2 \cup \mathbb M_3 )\cap (\mathbb N_1 \cup \mathbb N_2 \cup \mathbb N_3 )$ renders that the $K$ matrix possesses two boundary zero modes if the following condition is satisfied:
\begin{gather}
	\left\{\begin{array}{l}
        |m|<|l_2+l_3|,  \\
        l_4=0,\\
        n(l_2+l_3)= 2l_1m, \\
        l_2\neq l_3.
   \end{array}\right.
\end{gather}
In this case, the corresponding iCS field theory is a Type-I theory. Furthermore,
 the set operation $(\mathbb M_1 \cup \mathbb M_2 \cup \mathbb M_3 \cup \mathbb N_1 \cup \mathbb N_2 \cup \mathbb N_3 )\backslash [(\mathbb M_1 \cup \mathbb M_2 \cup \mathbb M_3 )\cap (\mathbb N_1 \cup \mathbb N_2 \cup \mathbb N_3 )]$ renders the following conditions for exactly one boundary zero mode:
\begin{gather}
	\left\{\begin{array}{l}
        |m|<|l_2+l_3|,  \\
        l_4=0,\\
        n(l_2+l_3)\neq 2l_1m,\\
        l_2\neq l_3 ,
   \end{array}\right.\text{ or }\left\{\begin{array}{l}
        |m|<|l_2-l_3|,  \\
        l_4\neq 0,\\
        \det B =l_1l_4-l_2l_3=0,\\
        (2l_2m-l_4n)(2l_3m-l_4n)=0.
   \end{array}\right.
\end{gather}
In this case, the corresponding iCS field theory is a Type-II iCS field theory. More specifically, if $\left\{\begin{array}{l}
l_4=0,\\ |l_2|<|l_3|,
\end{array}\right.$ or $\left\{\begin{array}{l}
l_4\neq 0,\\ 2l_3m=l_4n,
\end{array}\right.$ the $K$ matrix has a lower boundary zero mode satisfying the boundary condition Eq.~(\ref{Boundary11}); if $\left\{\begin{array}{l}
l_4=0,\\ |l_3|<|l_2|,
\end{array}\right.$ or $\left\{\begin{array}{l}
l_4\neq 0,\\ 2l_2m=l_4n,
\end{array}\right.$ the $K$ matrix has an upper boundary zero mode satisfying the boundary condition Eq.~(\ref{Boundary22}). The results are summarized in Table \ref{icstoeplitz}.
The aforementioned $K_1^{2N}$ matrix, in its mathematical form, is equivalent to the single-particle Hamiltonian of a typical 1d topological insulator i.e., the SSH model. Since $K_1^{2N}$ has two boundary zero modes, it belongs to Type-I iCS field theory. Generally speaking, Type-I $K$ matrix has been widely discussed in the study of TI and TSC. The presence of boundary zero mode, as is well-known, can be indicated by topological invariants. However, the Toeplitz matrices with only one boundary zero mode, i.e., the Type-II Toeplitz $K$ matrices, are rarely studied in the context of TI and TSC. Although we draw comparisons between the $K$ matrices of iCS field theories and the Hamiltonians of TI and TSC, it is important to note that the occurrence of Toeplitz braiding does not require any symmetry protection. 

In terms of the \textit{foliation structure}\cite{shirley2018fracton,ma2022fractonic,chen2023ground}, the Type-I and Type-II iCS field theories presented above are non-foliated. Within the context of iCS field theories, having a foliation structure means that one can apply $GL(N,\ZZ)$ transformation $K\rightarrow K'=W K W^\mT$ to extract a small block $K'$, i.e. $K=K'\oplus K''$, while leaving the Toeplitz structure of the matrix $K''$ unaltered. In order to preserve the locality in $z$-direction, $W\in GL(N,\ZZ)$ is equal to the identity except for a small diagonal block, corresponding to a local relabeling of gauge fields. Based on the necessary condition established in Ref.~\cite{chen2023ground}, a necessary condition for the foliation structure is that the determinant polynomial
\begin{align}
	D(u)&=\det(u^{-1}B+A+u B^\mT)\notag\\
	&=2l_1 l_4-l_2^2-l_3^2-m^2+(l_1l_4-l_2l_3)(\frac{1}{u^2}+u^2)\notag\\
	&+(l_4 n-l_2 m-l_3 m)(\frac{1}{u}+u)
\end{align}
must be a constant. For Type-I and Type-II iCS field theories, we can easily verify that $D(u)$ is not a constant. Therefore, Type-I and Type-II iCS field theories are non-foliated.

In order to demonstrate the nonlocal braiding statistics (Toeplitz braiding) along the $z$-direction in Type-I and Type-II iCS field theories, we present a numerical computation in the upcoming section.

\section{Numerical computation on iCS Field Theories with Toeplitz braiding\label{sectionIV}}

\begin{figure*}[htbp!]
    \centering
    \includegraphics[width=18cm]{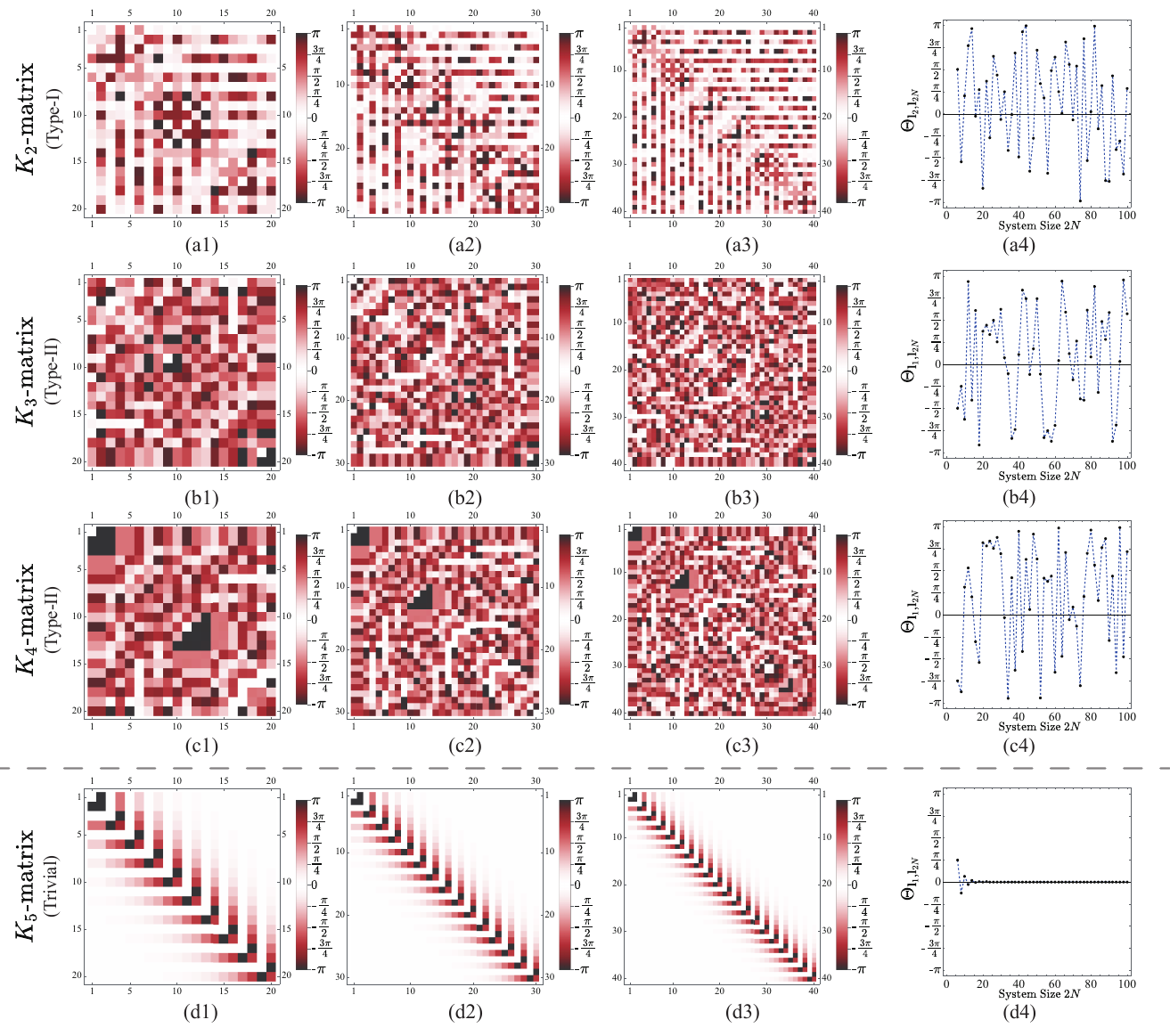}
    \caption{(a1-a3), (b1-b3) and (c1-c3) are the matrix plots of mutual braiding phase angle $\Theta_{\mathbf l_I, \mathbf l_{J}}=2 \pi \mathbf{l}_I^\mT (K_{2,3,4}^{2N})^{-1} \mathbf{l}_{J}$, where we take the system size $2N=20,30,40$ and the mutual braiding phase angle is limited to $[-\pi,\pi)$. (a4), (b4) and (c4) show that the mutual braiding phase angle between topological excitations residing at distinct $z$-boundaries exhibit oscillation as the system size $2N$ increases, where we choose $\Theta_{\bf l_2,\bf l_{2N}}$ or $\Theta_{\bf l_1,\bf l_{2N}}$ as a representative. (d4) demonstrates the braiding phase angle between topological excitations residing at distinct $z$-boundaries decreases drastically as the system size $2N$ increases in $K_5$-matrix (trivial) iCS field theory, where we select $\Theta_{\bf l_1,\bf l_{2N}}$ as a representative.}
    \label{bigfigure}
\end{figure*}

In Section \ref{icsnumerical1} and \ref{boundarymodebraiding}, we conduct a numerical computation that illustrates the connection between the appearance of nonlocal braiding statistics in $z$-direction and the existence of boundary zero modes of $K$ matrix. For comparative analysis, we focus on iCS field theories that arise from stacking $\mathbb{Z}_2$ twisted topological orders. The block $A$ in the $K$ matrix, given by (\ref{ktridiagonal}), is $\begin{pmatrix}
    2 & 2\\ 2& 0
\end{pmatrix}$, which describes the double-semion topological order. In Section \ref{fractionalizationc}, we present the symmetry fractionalization pattern in iCS field theories enriched by $U(1)$ charge conservation symmetry.  

\subsection{Numerical computation on Toeplitz braiding via Type-I and Type-II iCS field theories\label{icsnumerical1}}
To demonstrate the correspondence between boundary zero modes and Toeplitz braiding in Type-I and Type-II iCS field theories, we consider three iCS field theories as illustrative examples without loss of generality. The $K$ matrices (Eq.~(\ref{ktridiagonal})) are denoted as $K_2^{2N}$, $K_3^{2N}$ and $K_4^{2N}$, where $2N$  is the system size as well as the size of the $K$ matrix. 
Block $A$ and block $B$ in $K_2^{2N}$, $K_3^{2N}$ and $K_4^{2N}$ matrices are denoted as $A_j$ and $B_j$ ($j=2,3,4$), where
\begin{gather}
    A_2=A_3=A_4=\begin{pmatrix}
        2 & 2\\ 2& 0
    \end{pmatrix}
\end{gather}
and $B_j$ $(j=2,3,4)$ reads
\begin{gather}
B_2=\begin{pmatrix}
        2 & 1 \\ 3 & 0
    \end{pmatrix},\quad B_3=\begin{pmatrix}
        2 & 4 \\ 1 & 2
    \end{pmatrix},\quad B_4=\begin{pmatrix}
        2 & 1 \\ 4 & 2
    \end{pmatrix}.
\end{gather}
The elements of $K_2^{2N}$ satisfy the condition (\ref{kavakos2}) in Sec.~\ref{sectionIII}, implying that $K_2^{2N}$ has two boundary zero modes. Therefore, the corresponding iCS field theory is a Type-I iCS field theory.
The elements of $K_3^{2N}$ satisfy the condition (\ref{godel1}) , whereas $K_4^{2N}$ satisfy the condition (\ref{godel2}). Therefore, $K_3^{2N}$ possess one lower boundary zero mode while $K_4^{2N}$ exhibit one upper boundary zero modes. They belong to Type-II iCS field theories.
To demonstrate the nonlocal braiding statistics encoded in these iCS field theories, i.e., the \textit{Toeplitz braiding}, we numerically compute the braiding statistics encoded in finite-size $K$ matrices of different dimensions, from which how the braiding statistics scale with system size is known. 
The mutual braiding phase angle $\Theta_{\mathbf l_I,\mathbf l_J}$ is presented in Fig.~\ref{bigfigure}, where $\Theta_{\mathbf l_I,\mathbf l_J}$ (Eq.~(\ref{Boundary11})) is the mutual braiding phase angle between topological excitations carrying unit gauge charge (Eq.~(\ref{unitcharge})). Fig.~\ref{bigfigure}(a1-a3), Fig.~\ref{bigfigure}(b1-b3) and Fig.~\ref{bigfigure}(c1-c3) are the matrix plots of mutual braiding phase angle $\Theta_{\mathbf l_I,\mathbf l_J}$ of $K_2$-matrix, $K_3$-matrix and $K_4$-matrix iCS field theories, respectively, where we set the size of the $K$ matrix to be $20$, $30$ and $40$.  Numerical results show the robust presence of non-zero elements in the upper-right and lower-left positions of the matrices upon the system size is enlarged, which is vital to Toeplitz braiding. Fig.~\ref{bigfigure}(a4) shows that how the mutual braiding phase angle $\Theta_{\mathbf l_2,\mathbf l_{2N}}$ changes with respect to the system size $2N$, while Fig.~\ref{bigfigure}(b4) and Fig.~\ref{bigfigure}(c4) show how the mutual braiding phase angle $\Theta_{\mathbf l_1,\mathbf l_{2N}}$ changes with respect to the  system size $2N$. From the numerical results, we clearly find that  the phase angle oscillates as the system size $2N$ increases,  indicative of the existence of Toeplitz braiding. 

\begin{figure*}
    \centering
    \includegraphics[width=18cm]{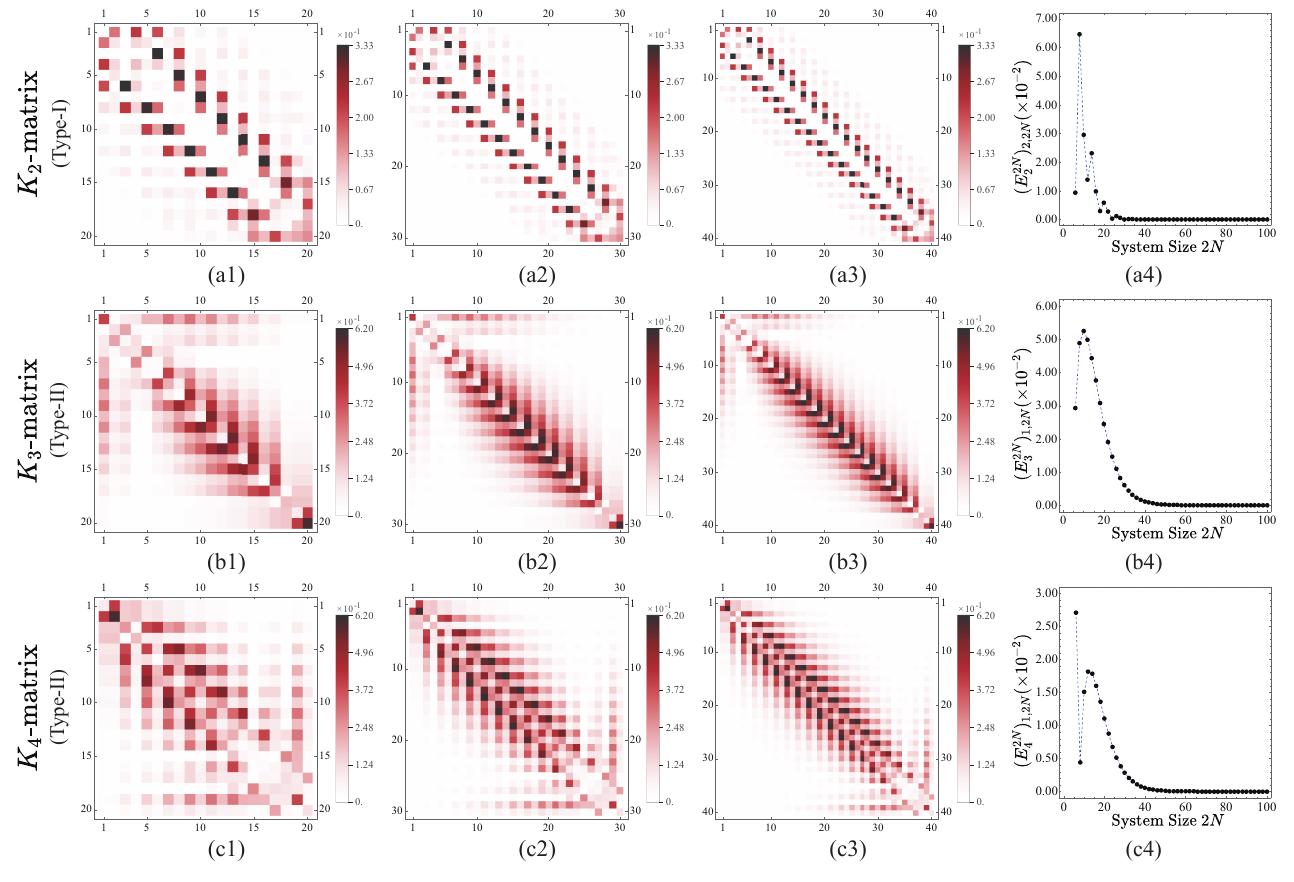}
    \caption{(a1-a3), (b1-b3) and (c1-c3) are the matrix plots of $E_2^{2N}$, $E_3^{2N}$ and $E_4^{2N}$, where we select 3 representative system size $2N=20,30,40$ for each type of iCS field theory. The presence of blank space in the upper-right and lower-left indicates that the matrices $M_2^{2N}$, $M_3^{2N}$ and $M_4^{2N}$ can effectively describe the braiding statistics between topological excitations residing at distinct $z$-boundaries. (a4), (b4) and (c4) shows that the upper-right and lower-left elements of $E_2^{2N}$, $E_3^{2N}$ and $E_4^{2N}$ decrease drastically as the system size $2N$ increases, where we choose $(E_2^{2N})_{2,2N}$, $(E_3^{2N})_{1,2N}$ and $(E_4^{2N})_{1,2N}$ as representatives.}
    \label{bigfigure2}
\end{figure*}

To more clearly understand the nontriviality of  Toeplitz braiding, it is beneficial to  make a sharp comparison by demonstrating  what we expect for the $K$ matrix that doesn't support Toeplitz braiding. For this purpose,  we  consider a new  $K$ matrix  denoted as $K_5^{2N}$, composed of the following $A$ and $B$ blocks
\begin{gather}
	 A_5=\begin{pmatrix}
        2 & 2\\ 2& 0
    \end{pmatrix},\quad B_5=\begin{pmatrix}
        0 & 1 \\ 0 & 0
    \end{pmatrix}.
\end{gather}
Fig.~\ref{bigfigure}(d1-d3) are the matrix plots of the mutual braiding phase angle $\Theta_{\mathbf l_I,\mathbf l_J}$ encoded in $K_5^{2N}$, where we also set the matrix size $2N=20,30,40$ during the numerical computation. The mutual braiding phase angle $\Theta_{\mathbf l_I,\mathbf l_J}$ decreases with $|I-J|$, i.e., the distance in $z$ between topological excitations. If two topological excitations are sufficiently far apart in $z$-direction, they cannot detect each other by braiding. Fig.~\ref{bigfigure}(d4) shows how the mutual braiding phase angle $\Theta_{\mathbf l_1,\mathbf l_{2N}}$ decreases drastically as the system size $2N$ increases, which also demonstrates that the topological excitations residing at distinct $z$-boundaries cannot detect each other via braiding.

\subsection{Numerical computation on the relation between Toeplitz braiding and boundary zero modes\label{boundarymodebraiding}}

In Sec.~\ref{sectionII}, we have shown that the braiding statistics between topological excitations at distinct $z$-boundaries is captured by boundary zero modes. Following this thought, we carry on a numerical computation, demonstrating that the correspondence still holds in  $K_2$-, $K_3$- and $K_4$-matrix iCS field theories. In analogy to  Eq.~(\ref{m12n}) in Sec.~\ref{sectionII}, for Type-I iCS field theories with two boundary zero modes $\mathbf v_1$ and $\mathbf v_2$, we claim that the braiding statistics between topological excitations residing at distinct $z$-boundaries are captured by
\begin{gather}
	M=\frac{1}{\lambda_1} \mathbf v_1 \mathbf v_1^\dag + \frac{1}{\lambda_2} \mathbf v_2 \mathbf v_2^\dag,
\end{gather}
where $\lambda_1$ and $\lambda_2$ are the corresponding exponentially suppressed eigenvalues of $\mathbf v_1$ and $\mathbf v_2$. As is discussed in Sec.~\ref{sectionIII}, for Type-II iCS field theories with only one boundary zero mode $\mathbf v$, we may also introduce a matrix $M$ constructed by the only boundary zero mode and the corresponding exponentially suppressed eigenvalue $\lambda$,
\begin{gather}
	M=\frac{1}{\lambda} \mathbf v \mathbf v^\dag.
\end{gather}
The matrix $M$ constructed by the boundary zero modes of $K_2^{2N}$, $K_3^{2N}$ and $K_4^{2N}$ are denoted as $M_2^{2N}$, $M_3^{2N}$ and $M_4^{2N}$, respectively. Compared to the inverse of the $K_2^{2N}$, $K_3^{2N}$ and $K_4^{2N}$, ${M_2^{2N}}$, $M_3^{2N}$ and $M_4^{2N}$ provide a good asymptotic approximation to the upper-right and lower-left elements, as depicted in Fig.~\ref{bigfigure2}(a1-a3), Fig.~\ref{bigfigure2}(b1-b3) and Fig.~\ref{bigfigure2}(c1-c3), where we define matrices $E_{j}^{2N}$ with element $(E_{j}^{2N})_{IJ}=|(M_{j}^{2N})_{IJ}-[(K_{j}^{2N})^{-1}]_{IJ}|$ ($j=2,3,4$) to characterize the difference between the two matrices $K_j^{-1}$ an $M_j$. To illustrate,  we choose $2N=20,30,40$ for each type of iCS field theories. As $2N$ increases, the upper-right and lower-left elements of $M_j^{2N}$ and $(K_j^{2N})^{-1}$ ($j=2,3,4$) converge towards each other, as shown in Fig.~\ref{bigfigure}(a4), Fig.~\ref{bigfigure}(b4) and Fig.~\ref{bigfigure}(c4), where we focus on the asymptotic behavior of $(E_2^{2N})_{2,2N}$, $(E_3^{2N})_{1,2N}$ and $(E_4^{2N})_{1,2N}$ as representatives.

This observation suggests that we can extract information about the braiding statistics of excitations residing at distinct $z$-boundaries by analyzing the zero modes and their ``close-to-zero'' eigenvalues.

\subsection{Charge fractionalization in iCS field theory\label{fractionalizationc}}

The discussion on Toeplitz braiding primarily addresses the impact of the boundary zero modes of $K$ on boundary physics. The influence of these boundary zero modes on $z$-bulk physics is manifested in the iCS field theory enriched by $U(1)$ charge-conservation symmetry. This subsection explores such symmetry enrichment and provides the results of numerical computation on charge fractionalization of topological excitations. Compared to the iCS field theories with periodic boundary condition (PBC) in the $z$-direction, we find that the pure mathematical entities-boundary zero modes of $K$, indeed influence the symmetry charges carried by topological excitations residing at $z$-bulk. 

\begin{figure}
    \centering
    \includegraphics[width=9cm]{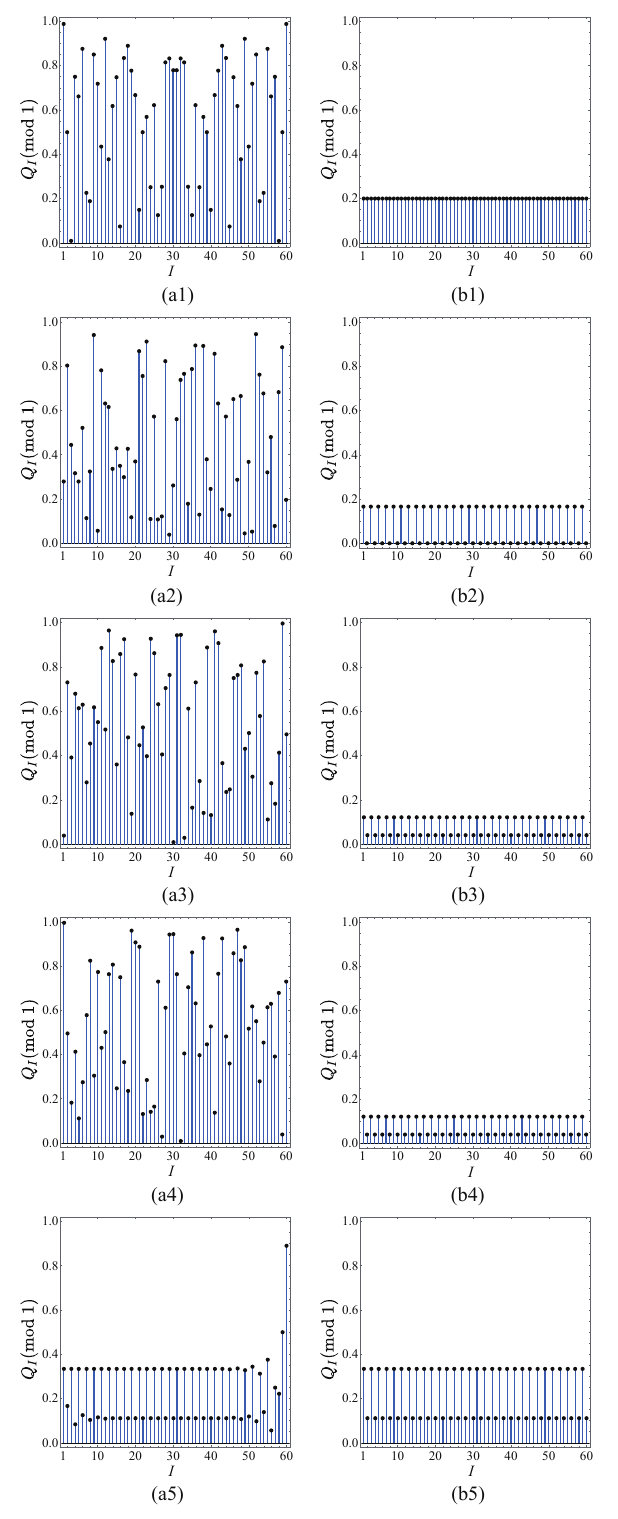}
    \caption{(a1-a5) are the fractionalized charges $Q_I$ carried by topological excitations $\mathbf l_I$ in $K_{1,2,3,4,5}$-matrix iCS field theories with OBC applied in $z$-direction. In comparison, (b1-b5) are the fractionalized charges $Q_I$ of topological excitations with PBC applied in $z$-direction. For illustrative purpose, we select the system size $2N=60$.}
    \label{icsfractionalization}
\end{figure}

 In the field theoretical formalism, symmetry charges are then carried by the currents $J^{I,\mu}=\frac{1}{2\pi}\epsilon^{\mu\nu\lambda}\partial_\nu a_\lambda^I$. In order to probe the conserved currents $J^{I,\mu}$, each $J^{I,\mu}$ is minimally coupled to an external electromagnetic field $A_\mu$. Consequently, in addition to the pure Chern-Simons term (Eq.~(\ref{lagrangianmcs})) and the excitation term, we incorporate an additional term $\sum_{I=1}^{2N} \frac{q^I}{2\pi} A_\mu  \epsilon^{\mu\nu\lambda}\partial_\nu a_\lambda^I$ into the Lagrangian\cite{wen2004quantum}. $q^I\in\ZZ$ composes a so-called charge vector $\mathbf q=(q^1,\ldots,q^{2N})^\mT$. By integrating $a_\mu^I$ out, we obtain the effective $A_\mu$ charge $Q_I$ carried by the topological excitation $\mathbf l_I$ (Eq.~(\ref{unitcharge})).  
\begin{gather}
	Q_I=\mathbf q K^{-1} \mathbf l_I.
\end{gather}
Here, we select $q^I=1$ for $I=1,\ldots,2N$, which implies the trivial excitations carry integral $A_\mu$ charge. As a result, in the study of symmetry fractionalization, the key aspect is $Q_I\mod 1$, as the integral part of $Q_I$ does not matter. In the following discussion, we drop the integral part of $Q_I$. The effective $A_\mu$ charges $Q_I$ carried by $\mathbf l_I$ ($I=1,\ldots,2N$) in $K_1$(with $m=2$, $l=3$), $K_2$, $K_3$ and $K_4$-matrix iCS field theory are plotted in Fig.~\ref{icsfractionalization}(a1-a4). For illustrative purposes, we set the system size $2N=60$. $Q_I$ exhibits random oscillations as $I$ varies. To make comparisons, we plot the effective $A_\mu$ charge $Q_I$ of topological excitations under PBC in the $z$-direction in Fig.~\ref{icsfractionalization}(b1-b4). As is observed, $Q_I$ displays a translational symmetry, such that $Q_I=Q_{I+2}$. The comparison demonstrates that the random oscillation of $Q_I$ is due to the influence of boundary zero modes of $K$.
 
 In contrast, for the (gapped) iCS field theories without Toeplitz braiding, even with OBC in $z$, $Q_I$ exhibits translational symmetry for $I$ corresponding to the $z$-bulk coordinates ($1\ll I\ll 2N$), such that $Q_I\simeq Q_{I+2}$, consistent with the charge fractionalization pattern observed under PBC. In Fig.~\ref{icsfractionalization}(a5) and Fig.~\ref{icsfractionalization}(b5), we present the fractionalized charges $Q_I$ of topological excitations in $K_5$-matrix iCS field theories. Fig.~\ref{icsfractionalization}(a5) is under OBC while Fig.~\ref{icsfractionalization}(b5) is under PBC.  From this, we understand that in the absence of boundary zero modes of the $K$ matrix, (gapped) iCS field theories with PBC provide a good description to the $z$-bulk physics in OBC.

\section{Conclusions\label{sectionV}}

In this paper, we constructed a type of 3D topological fracton topological orders  whose surface states support  a nontrivial form of braiding statistics, termed \textit{Toeplitz braiding}. This boundary phenomenon   arises due to the block-tridiagonal Toeplitz structure of the \(K\)-matrix in the infinite-component Chern-Simons theory. This phenomenon represents a significant departure from conventional braiding statistics observed in 2D topological orders. Here, we  further  summarize the paper in details. First,   in the field of  free-fermion topological insulators, boundary zero modes of Toeplitz matrices play a vital role. In this paper, we  substantially extend the topological physics of boundary zero modes of Toeplitz matrices to the present strongly correlated systems. Second, our research categorizes nontrivial \(K\)-matrices into two distinct types based on their boundary zero modes. This classification provides a deeper understanding of the underlying mathematical structures and their physical implications in 3D fracton topological orders. Third, we conduct extensive numerical simulations and finite-size scaling analyses to confirm the theoretical predictions. These simulations demonstrate the robustness of Toeplitz braiding. Fourth, we also provide a discussion on symmetry fractionalization in iCS field theories, concluding that the boundary zero modes of $K$ influence the fractionalization pattern. 

  There are several future directions. 
First of all, the categorization of iCS field theories with Toeplitz braiding is primarily for mathematical convenience. How to physically distinguish Type-I and Type-II iCS field theories is left for future research.
It will also be interesting to construct the lattice model realization of iCS field theories that supports boundary Toeplitz braiding, which may pave the way for further numerical or experimental study on Toeplitz braiding. From the perspective of fractional quantum Hall states or  spin liquid, the stacking process described in Sec.~\ref{sectionIIB} can be realized  by condensing topological excitations followed by adding interlayer interaction of condensation currents or spin-spin interaction~\cite{blok1990effective,blok1990hierarchy,wen1992classification,wu2023transition}, which may be also useful here.  Detailed construction that explicitly enables Toeplitz braiding will be left to future discussion. 
We might also consider enlarging blocks $A$ and $B$ in the Toeplitz $K$ matrix, which might affect the number of boundary zero modes and reshape the braiding statistics. Breaking the strict Toeplitz character of the $K$ matrix and considering the ``stability" of braiding statistics against stacking faults is also an intriguing topic.
Furthermore, following the construction of topologically-ordered non-liquid states in this paper, the construction of higher-dimensional topologically-ordered non-liquid states is directly allowed. For instance, by stacking (3+1)D topological orders~\cite{zhang2021compatible,zhang2023continuum,zhang2023non,2024arXiv240519077H}, we can obtain a (4+1)D topologically-ordered non-liquid state. Similarly, stacking (2+1)D topological orders in two independent spatial dimensions can also yield (4+1)D topologically-ordered non-liquid states. The study of these theories and the duality between them is left for future research.  We can also compute entanglement entropy via partitioning the system into two subsystems along $z$-direction and study whether and how the presence of Toeplitz braiding leaves its fingerprint in the scaling behavior of entanglement entropy.

\acknowledgments
This work was supported by National Natural Science Foundation of China (NSFC) Grant No. 12074438 and Guangdong Provincial Key Laboratory of Magnetoelectric Physics and Devices under Grant No. 2022B1212010008.





\onecolumngrid

\appendix

\section{Detailed Derivation on Sorting Boundary Zero Modes of Toeplitz $K$ matrices\label{zeromodes}}

\begin{figure*}
\flushleft
    \begin{tikzpicture}[
  box/.style={draw, rectangle, rounded corners, minimum width=1cm, minimum height=0.5cm, text centered, fill=white!20},
  line/.style={draw}
]

  \node[box] (main) {$\begin{array}{c}K\mathbf u=0 \\(N\rightarrow \infty)\end{array}$};

  \node[box, above right=2cm and 0.5cm of main] (sub1) {$\det B=0$};
  \node[box, below right=2cm and 0.5cm of main] (sub2) {$\begin{array}{c}
  	\det B\neq 0,\\ l_4= 0
  \end{array}$};
  \node[box, below=2cm of sub2] (exception) {$\begin{array}{c}
  	\det B\neq 0,\ l_4\neq 0:\\
  	\text{case-by-case study}
  \end{array}$ };

  \node[box, above right=1cm and 0.8cm of sub1] (sub11)  {Eq.~(\ref{Boundary1})};
  \node[box, below right=1cm and 0.8cm of sub1] (sub12) {Eq.~(\ref{Boundary2})};
  \node[above=1cm of sub11] (subtitle1) {$\begin{array}{c}
  	\text{Boundary condition} \\ \text{being discussed}
  \end{array}$};
  \node[box, above right=0.7cm and 0.5cm of sub2] (sub21)  {Eq.~(\ref{Boundary1})};
  \node[box, below right=0.7cm and 0.5cm of sub2] (sub22) {Eq.~(\ref{Boundary2})};
 
  \node[box, above right=0.2cm and 0.5cm of sub11] (sub111)  {$\begin{array}{c}\text{Discussion on}\\ \text{$l_4\neq 0$ cases}\end{array}$};
  \node[box, below right=0.2cm and 0.5cm of sub11] (sub112)  {$\begin{array}{c}\text{Discussion on}\\ \text{$l_4= 0$ cases}\end{array}$};
 \node[box, above right=0.2cm and 0.5cm of sub12] (sub121)  {$\begin{array}{c}\text{Discussion on}\\ \text{$l_4\neq 0$ cases}\end{array}$};
  \node[box, below right=0.2cm and 0.5cm of sub12] (sub122)  {$\begin{array}{c}\text{Discussion on}\\ \text{$l_4 = 0$ cases}\end{array}$};
  
 \node[box, right=0.5cm of sub21] (sub211)  {$\begin{array}{c}\text{Discussion}\\ \text{on Case I $\sim$}\\ \text{Case IV}\end{array}$};
 \node[box, right=0.5cm of sub22] (sub221)  {$\begin{array}{c}\text{Discussion}\\ \text{on Case I $\sim$}\\ \text{Case IV}\end{array}$};

\node[box, right=3.1cm of sub11] (sub1111)  {$\begin{array}{c}
	\text{Conditions for a}\\ \text{boundary zero} \\ \text{mode:} \\ \text{Eq.~(\ref{sushi1})}
\end{array}$};
\node[box, right=3.1cm of sub12] (sub1211)  {$\begin{array}{c}
	\text{Conditions for a}\\ \text{boundary zero} \\ \text{mode:} \\ \text{Eq.~(\ref{sushi2})}
\end{array}$};
\node[box, right=0.61cm of sub211] (sub2111)  {$\begin{array}{c}
	\text{Conditions for}\\ \text{a boundary zero} \\ \text{mode:}\\ \text{Eq.~(\ref{libai1})}
\end{array}$};
\node[box, right=0.61cm of sub221] (sub2211)  {$\begin{array}{c}
	\text{Conditions for a}\\ \text{boundary zero} \\ \text{mode:}\\ \text{Eq.~(\ref{libai2})}
\end{array}$};

\node[box, right=8.3cm of sub1] (sub11111) {$\begin{array}{c}
	\text{Conditions for a} \\ \text{boundary zero} \\ \text{mode satisfying} \\ \text{Eq.(\ref{Boundary1}): Eq.~(\ref{dufu1})}
\end{array}$};
\node[box, right=8cm of sub2] (sub12111) {$\begin{array}{c}
	\text{Conditions for a} \\ \text{boundary zero} \\ \text{mode satisfying} \\ \text{Eq.(\ref{Boundary2}): Eq.~(\ref{dufu2})}
\end{array}$};

\node[box, right=13.7cm of main] (sub111111) {$\begin{array}{c}
	\text{Type-I and Type-II} \\ \text{$K$ matrices} \\ \text{(Table \ref{icstoeplitz})}
\end{array}$};

  \draw[line] (main.east) -- (sub1.west);
  \draw[line] (main.east) -- (sub2.west);
  
  \draw[line] (sub1.east) -- (sub11.west);
  \draw[line] (sub1.east) -- (sub12.west);
  \draw[line] (sub2.east) -- (sub21.west);
  \draw[line] (sub2.east) -- (sub22.west);
  \draw[line] (sub2.south) -- (exception.north);
  
  \draw[line] (sub11.east) -- (sub111.west);
  \draw[line] (sub11.east) -- (sub112.west);
  \draw[line] (sub12.east) -- (sub121.west);
  \draw[line] (sub12.east) -- (sub122.west);
  \draw[line] (sub21.east) -- (sub211.west);
  \draw[line] (sub22.east) -- (sub221.west);
  
  \draw[line] (sub111.east) -- (sub1111.west);
  \draw[line] (sub112.east) -- (sub1111.west);
  \draw[line] (sub121.east) -- (sub1211.west);
  \draw[line] (sub122.east) -- (sub1211.west);
  \draw[line] (sub211.east) -- (sub2111.west);
  \draw[line] (sub221.east) -- (sub2211.west);
  
  \draw[line] (sub1111.east) -- (sub11111.west);
  \draw[line] (sub2111.east) -- (sub11111.west);
  \draw[line] (sub1211.east) -- (sub12111.west);
  \draw[line] (sub2211.east) -- (sub12111.west);
  
  \draw[line] (sub11111.east) -- (sub111111.west);
  \draw[line] (sub12111.east) -- (sub111111.west);
  
\end{tikzpicture}
\caption{
Mind map for sorting boundary zero modes of the $K$ matrix (Eq.~(\ref{ktridiagonal})). We categorize our discussion based on whether $\det B$ is zero or not, which affects the \# of boundary bases as well as the structure of trial solutions. The discussion on $\det B=0$ cases is presented in Appendix \ref{section1}, where we divide our discussion into Subsection \ref{goddamn1} and Subsection \ref{goddamn2}, addressing boundary conditions Eq.~(\ref{Boundary1}) and Eq.~(\ref{Boundary2}), respectively. Within each subsection, the discussion is organized into two distinct cases, based on whether $l_4$ is zero or not. After a thorough discussion, the conditions that yield a boundary zero mode satisfying the boundary condition Eq.~(\ref{Boundary1}) are summarized in Eq.~(\ref{sushi1}), while the conditions for a boundary zero mode satisfying the boundary condition Eq.~(\ref{Boundary2}) are summarized in Eq.~(\ref{sushi2}).
The discussion on $\det B\neq 0$ $K$ matrices is presented in Appendix \ref{section2}. There, we provide a detailed analysis specifically for $K$ matrices with $\det B\neq 0$, $l_4=0$. Similar to the discussion on $K$ matrices with $\det B=0$, we divide our discussion into two subsections: Subsection \ref{chewinggum1} and \ref{chewinggum2}, addressing distinct boundary conditions  Eq.~(\ref{Boundary1}) and Eq.~(\ref{Boundary2}), respectively. In each subsection, the discussion is divided into four cases: Case I $\sim$ Case IV. The conditions for boundary zero modes are summarized in Eq.~(\ref{libai1}) and Eq.~(\ref{libai2}). Taking the union of Eq.~(\ref{sushi1}) and Eq.~(\ref{libai1}), we obtain the conditions for appearance of a boundary zero mode satisfying the boundary condition Eq.~(\ref{Boundary1}). Likewise, taking the union of Eq.~(\ref{sushi2}) and Eq.~(\ref{libai2}), we obtain the conditions for appearance of a boundary zero mode satisfying the boundary condition Eq.~(\ref{Boundary2}). Set operations render the conditions for only one or two boundary zero modes, i.e. the condition for Type-I and Type-II $K$ matrices in Table \ref{icstoeplitz}.
The analysis on $K$ matrices with $\det B\neq 0$, $l_4\neq 0$ is more challenging, requiring a case-by case study, and is left for future research.
}
    \label{mindmap}
\end{figure*}

The Appendix provides further discussions on the technical details of sorting boundary zero modes of the $K$ matrix (Eq.~(\ref{ktridiagonal})) in Sec.~\ref{sectionIII} of the main text. 
To facilitate reading, we outline the sorting procedures below, and list key equations.
As is described in Sec.~\ref{sectionIII}, the ansatz for a trial solution
$\mathbf{u}$ of boundary zero mode satisfying $K\mathbf u=0$ in the thermodynamic limit $N\rightarrow \infty$ is a linear combination of boundary bases $\{\mathbf{w}^j\}$:
\begin{gather}
    \mathbf{u}=\sum_j c_j \mathbf w^{j}.\label{lifesucks1}
\end{gather}
Here the boundary base $\mathbf w^j$ has the form
\begin{gather}
    \mathbf w^j=\begin{pmatrix}
         \beta_jw^{j,a} & \beta_jw^{j,b} & \beta_j^2 w^{j,a} & \beta_j^2 w^{j,b} & \cdots & \beta_j^{N} w^{j,a} & \beta_j^{N} w^{j,b}
    \end{pmatrix}^\mT,\label{phij}
\end{gather}
which is required to satisfy the bulk recurrence relation
\begin{gather}
    B \begin{pmatrix}
        w^{j,a} \\ w^{j,b}
    \end{pmatrix}+ \beta_j A\begin{pmatrix}
        w^{j,a} \\ w^{j,b}
    \end{pmatrix}+\beta_j^2 B^\mT \begin{pmatrix}
        w^{j,a} \\ w^{j,b}
    \end{pmatrix}=0.\label{Bulk1}
\end{gather}
Nontrivial solution to $\begin{pmatrix}
    w^{j,a} & w^{j,b}
\end{pmatrix}^\mT$ renders the equation for $\beta_j$
\begin{gather}
    \det(B+A\beta_j+B^\mT\beta_j^2)=0.\label{beijing1}
\end{gather}
Therefore, $\{\beta_j\}$ are the solutions to the equation
\begin{gather}
\alpha_1+\alpha_2\beta+\alpha_3\beta^2+\alpha_2\beta^3+\alpha_1\beta^4=0 , \label{equationforbeta11} \\ \left\{
\begin{array}{l}
     \alpha_1=l_1l_4-l_2l_3, \\ \alpha_2=-(m(l_2+l_3)-nl_4),\\
     \alpha_3=-(l_2^2+l_3^2-2l_1l_4+m^2).
\end{array}\right.
\end{gather}
The solutions $\{\beta_j\}$ to Eq.~(\ref{equationforbeta11}) together with the bulk recurrence relation Eq.~(\ref{Bulk1}) settle down the boundary bases $\{\mathbf w^j\}$ in Eq.~(\ref{phij}) (up to a normalization factor). The boundary bases are utilized to construct trial solutions.
 It is required that a trial solution $\mathbf{u}$ localizes at only one $z$-boundary, i.e., all the  $\{\beta_j\}$ in one trial solution $\mathbf{u}$ satisfies either $|\beta_j|>1$ or $|\beta_j|<1$.
If a trial solution $\mathbf u$ is indeed a legitimate boundary zero mode consisting of boundary bases $\{\mathbf w^j\}$ with $|\beta_j|<1$, it satisfies the boundary condition
\begin{subequations}
\begin{equation}
        A\begin{pmatrix}
        u_1^{a} \\ u_1^{b}
    \end{pmatrix}+B^\mT \begin{pmatrix}
        u_{2}^{a} \\ u_{2}^{b}
    \end{pmatrix}=0.\label{Boundary1}
    \end{equation}
Likewise, if a trial solution $\mathbf u$ is a legitimate boundary zero mode consisting of boundary bases $\{\mathbf w^j\}$ with $|\beta_j|>1$, it satisfies the boundary condition
\begin{equation}
         B \begin{pmatrix}
        u_{N-1}^{a} \\ u_{N-1}^{b}
    \end{pmatrix}+A\begin{pmatrix}
        u_N^{a} \\ u_N^{b}
    \end{pmatrix}=0.\label{Boundary2}
\end{equation}
\end{subequations}
Therefore, we can identify the conditions on the elements of the $K$ matrix that allow trial solutions to satisfy the boundary condition in Eq.~(\ref{Boundary1}) or Eq.~(\ref{Boundary2}), i.e. the conditions that enable boundary zero modes. 
As is discussed in the main text, we classify our discussion based on the degree of Eq.~(\ref{equationforbeta11}), which affects the \# of boundary bases as well as the structure of trial solutions. More concretely, the degree of Eq.~(\ref{equationforbeta11}) is determined by $\alpha_1=l_1l_4-l_2l_3=\det B$. 

If $\det B= 0$, Eq.~(\ref{equationforbeta11}) reduces to a cubic equation. 
In this case, one solution is $\beta_0=0$, which is discarded, leaving the two remaining solutions $\beta_1$ and $\beta_2$. They are determined by the quadratic equation
\begin{gather}
    \alpha_2+\alpha_3\beta_{1,2}+\alpha_2\beta_{1,2}^2=0,\label{damn1}
\end{gather}
from which $\beta_1\beta_2=1$ is immediately apparent.
Here we require $\beta_1$ satisfies $0<|\beta_1|<1$, and the corresponding boundary base is denoted as $\mathbf w^1$. The other solution $\beta_2$ satisfies $|\beta_2|>1$, of which the corresponding boundary base is denoted as $\mathbf w^2$. 
In Appendix \ref{section1}, we provide detailed discussions on the conditions for the elements of the $K$ matrix that enable trial solutions $\mathbf u^1=\mathbf w^1$  or $\mathbf u^2=\mathbf w^2$ being legitimate boundary zero modes. The discussion on $\mathbf u^1$ satisfying the lower boundary condition Eq.~(\ref{Boundary1}) is presented  in Subsection \ref{goddamn1}, while the discussion on $\mathbf u^2$ satisfying the upper boundary condition Eq.~(\ref{Boundary2}) is presented in Subsection \ref{goddamn2}.
 Furthermore, the discussion in Subsection \ref{goddamn1} or \ref{goddamn2} is categorized into $l_4\neq 0$ and $l_4=0$, which is due to the explicit form of $\mathbf w^{1,2}$ being different in these cases. 
Finally, the condition for the appearance of boundary zero modes under the $\det B=0$ condition  are summarized in Subsection \ref{goddamn3}.
We discover that the following set of conditions yields a boundary zero mode satisfying the boundary condition Eq.~(\ref{Boundary1}):
\begin{gather}
	\mathbb A_1: \left\{\begin{array}{l}
    m\neq 0,\quad l_4\neq 0,\quad l_2\neq l_3,\\
    l_1l_4=l_2l_3,\quad
    2l_3m=l_4n,\\
    |m|<|l_3-l_2|,
    \end{array}\right. \quad \text{or}\quad \mathbb A_2: \left\{\begin{array}{l}
    m\neq 0,\quad
    l_3\neq 0,\\
    l_4=l_2= 0,\\
         |m|<|l_3|,
    \end{array}\right. \quad \text{or} \quad  \mathbb A_3: \left\{\begin{array}{l}
    m\neq 0,\quad l_2\neq 0,\\
    nl_2=2l_1m,\quad
    l_3=l_4=0,\\
    |m|<|l_2|.
    \end{array}\right.\label{sushi1}
\end{gather}
Similarly, the following set of conditions yields a boundary zero mode satisfying the boundary condition Eq.~(\ref{Boundary2}):
\begin{gather}
	\mathbb B_1: \left\{\begin{array}{l}
    m\neq 0,\quad 
    l_4\neq 0,\quad l_2\neq l_3,\\
    l_1l_4=l_2l_3,\quad
    2l_2m=l_4n,\\
    |m|<|l_3-l_2|,
    \end{array}\right. 
    \quad \text{or} \quad \mathbb B_2:
    \left\{\begin{array}{l}
    m\neq 0,\quad 
    l_2\neq 0,\\
    l_4=l_3= 0,\\
         |m|<|l_2|,
    \end{array}\right. 
    \quad \text{or} \quad \mathbb B_3:
    \left\{\begin{array}{l}
    m\neq 0,\quad l_3\neq 0,\\
    nl_3=2l_1m,\quad
    l_2=l_4=0,\\
    |m|<|l_3|.
    \end{array}\right.\label{sushi2}
\end{gather}

If $\det B\neq 0$, Eq.~(\ref{equationforbeta11}) is a fourth-degree equation. In this case, if the element $l_4$ in block $B$ is nonzero, carrying on a comprehensive analytical analysis is challenging and requires a case-by-case study. To simplify discussion, we focus on the special case where $l_4=0$. 
In this situation, the equation for $\{\beta_j\}$ (Eq.~(\ref{equationforbeta11})) factorizes into 
\begin{gather}
	(l_2+m\beta_j+l_3\beta_j^2)(l_3+m\beta_j+l_2\beta_j^2)=0,\quad j=1,2,3,4,
\end{gather}
which facilitates our discussion. The detailed discussion is presented in Appendix~\ref{section2}, where we divide our discussion into Subsection \ref{chewinggum1} and Subsection \ref{chewinggum2}, addressing boundary conditions Eq.~(\ref{Boundary1}) and Eq.~(\ref{Boundary2}), respectively.
Within each subsection, the discussion is organized into four distinct cases: Case I $\sim$ Case IV. 
For each case, trial solutions are constructed using the boundary bases. 
Each trial solution, denoted as $\mathbf u^{i,j}$, is a linear superposition of two boundary bases, $\mathbf w^i$ and $\mathbf w^j$ (Eq.~(\ref{phij})), where the moduli of parameters $\beta_i$ and $\beta_j$ are either both greater than 1 or both less than 1. These trial solutions are required to match the corresponding boundary conditions.
This procedure yields explicit constraints on the elements of the $K$ matrix that guarantee the existence of boundary zero modes.
We discover that the following conditions yield a boundary zero mode satisfying the boundary condition Eq.~(\ref{Boundary1}):
\begin{gather}
	\mathbb C_1: \left\{
\begin{array}{l}
l_4=0,\quad n(l_2+l_3)=2l_1 m,\\
m\neq 0,\quad  l_2l_3\neq 0, \\
|m|<|l_2+l_3|,\quad |l_3|<|l_2|,\\
\end{array}
\right.\quad \text{or} \quad  \mathbb C_2: 
\left\{\begin{array}{l}
	    l_4=0,\\
		m\neq 0,\quad  l_2l_3\neq 0,\\
		|m|<|l_2+l_3|,\quad |l_2|<|l_3|.
	\end{array}\right.\label{libai1}
\end{gather}
Similarly, the following conditions yield a boundary zero mode satisfying the boundary condition Eq.~(\ref{Boundary2}):
\begin{gather}
	\mathbb D_1: \left\{\begin{array}{l}
	    l_4=0,\\
		m\neq 0,\quad  l_2l_3\neq 0,\\
		|m|<|l_2+l_3|,\quad |l_3|<|l_2|,
	\end{array}\right. \quad \text{or} \quad  \mathbb D_2: 
	\left\{
\begin{array}{l}
l_4=0,\quad n(l_2+l_3)=2l_1 m,\\
m\neq 0,\quad  l_2l_3\neq 0, \\
|m|<|l_2+l_3|,\quad |l_2|<|l_3|.\\
\end{array}
\right.\label{libai2}
\end{gather}

Taking the union of (\ref{sushi1}) and (\ref{libai1}), we find that the $K$ matrix Eq.~(\ref{ktridiagonal}) has a boundary zero mode matching the boundary condition Eq.~(\ref{Boundary1}) if one of the following conditions is satisfied:
\begin{gather}
	\mathbb M_1: \left\{\begin{array}{l}
    m\neq 0,\quad l_4\neq 0,\\
    l_1l_4=l_2l_3,\quad
    2l_3m=l_4n,\\
    |m|<|l_3-l_2|,
    \end{array}\right. 
    \quad \text{or} \quad \mathbb M_2:
    \left\{
\begin{array}{l}
m\neq 0, \\
l_4=0,\quad n(l_2+l_3)=2l_1 m,\\
|m|<|l_2+l_3|,\quad |l_3|<|l_2|,\\
\end{array}
\right.\quad \text{or} \quad \mathbb M_3:
\left\{\begin{array}{l}
		m\neq 0,\\
		l_4=0,\\
		|m|<|l_2+l_3|,\quad |l_2|<|l_3|.
	\end{array}\right.\label{dufu1}
\end{gather}
Similarly, taking the union of (\ref{sushi2}) and (\ref{libai2}), we find that the $K$ matrix Eq.~(\ref{ktridiagonal}) has a boundary zero mode matching the boundary condition Eq.~(\ref{Boundary2}) if one of the following  conditions is satisfied:
\begin{gather}
	\mathbb N_1: \left\{\begin{array}{l}
    m\neq 0,\quad l_4\neq 0,\\
    l_1l_4=l_2l_3,\quad
    2l_2m=l_4n,\\
    |m|<|l_3-l_2|,
    \end{array}\right. 
    \quad \text{or} \quad \mathbb N_2: 
    \left\{
\begin{array}{l}
m\neq 0, \\
l_4=0,\quad n(l_2+l_3)=2l_1 m,\\
|m|<|l_2+l_3|,\quad |l_2|<|l_3|,\\
\end{array}
\right.\quad \text{or} \quad \mathbb N_3: 
\left\{\begin{array}{l}
		m\neq 0,\\
		l_4=0,\\
		|m|<|l_2+l_3|,\quad |l_3|<|l_2|.
	\end{array}\right.\label{dufu2}
\end{gather}

The condition for the $K$ matrix (Eq.~(\ref{ktridiagonal})) possessing two boundary zero modes, i.e. being a Type-I $K$ matrix, is obtained by the set operation $(\mathbb M_1 \cup \mathbb M_2 \cup \mathbb M_3)  \cap(\mathbb N_1 \cup \mathbb N_2 \cup \mathbb N_3)$. The result is shown in the first row of Table \ref{icstoeplitz}.
\begin{gather}
	\left\{\begin{array}{l}
	m\neq 0,\quad         l_2\neq l_3,\\
        |m|<|l_2+l_3|,  \\
        l_4=0,\quad
        n(l_2+l_3)= 2l_1m.
   \end{array}\right.
\end{gather}

Furthermore, set operation $(\mathbb M_1 \cup \mathbb M_2 \cup \mathbb M_3\cup \mathbb N_1 \cup \mathbb N_2 \cup \mathbb N_3)\backslash [(\mathbb M_1 \cup \mathbb M_2 \cup \mathbb M_3)  \cap(\mathbb N_1 \cup \mathbb N_2 \cup \mathbb N_3)]$ yields that the $K$ matrix Eq.~(\ref{ktridiagonal}) has only one boundary zero mode, i.e. a Type-II $K$ matrix, if the elements in blocks $A$ and $B$ satisfy one of the following conditions
\begin{gather}
	\left\{\begin{array}{l}
        m\neq 0,\quad l_2\neq l_3,\\
        |m|<|l_2+l_3|,  \\
        l_4=0,\\
        n(l_2+l_3)\neq 2l_1m,
   \end{array}\right.\text{ or }\left\{\begin{array}{l}
   m\neq 0,\quad l_4\neq 0,\\
        |m|<|l_2-l_3|,  \\
        \det B =l_1l_4-l_2l_3=0,\\
        (2l_2m-l_4n)(2l_3m-l_4n)=0.
   \end{array}\right.
\end{gather}
More specifically, $\left\{\begin{array}{l}
	l_4=0\\
	|l_2|<|l_3|
\end{array}\right.$ or $\left\{\begin{array}{l}
	l_4\neq 0\\
	2l_3 m=l_4 n
\end{array}\right.$ indicates that the boundary zero mode is a lower boundary zero mode satisfying the boundary condition Eq.~(\ref{Boundary1}); $\left\{\begin{array}{l}
	l_4=0\\
	|l_3|<|l_2|
\end{array}\right.$ or $\left\{\begin{array}{l}
	l_4\neq 0\\
	2l_2 m=l_4 n
\end{array}\right.$ indicates that the boundary zero mode is an upper boundary zero mode satisfying the boundary condition Eq.~(\ref{Boundary2}). The results are summarized in the second row of Table \ref{icstoeplitz}.

In order to streamline the lengthy discussion, a mind map is presented in Fig.\ref{mindmap}.

\section{The Conditions for the Existence of Boundary Zero Modes under $\det B$=0\label{section1}}
Under the condition $\det B=0$, the equation for $\beta_j$ (Eq.~(\ref{equationforbeta11})) reduces to a quadratic equation Eq.~(\ref{damn1}), from which $\beta_1\beta_2=1$ is immediately apparent.
Here we require $\beta_1$ satisfies $0<|\beta_1|<1$, and the corresponding boundary base is denoted as $\mathbf w^1$. The other solution $\beta_2$ satisfies $|\beta_2|>1$, of which the corresponding boundary base is denoted as $\mathbf w^2$. Each trial solution consists of only one boundary base, i.e. $\mathbf u^1=\mathbf w^1$ and $\mathbf u^2=\mathbf w^2$.
In this subsection, we are going to figure out the explicit conditions that enable boundary zero modes. The discussion on $\mathbf u^1$ and $\mathbf u^2$ are presented separately in Subsection \ref{goddamn1}
 and \ref{goddamn2}.
 Furthermore, the discussions in Subsection \ref{goddamn1} and \ref{goddamn2} are categorized into $l_4\neq 0$ and $l_4=0$, which is due to the explicit form of $\mathbf w^{1,2}$ being different in these cases. 
 The conditions for the appearance of boundary zero modes under $\det B=0$  are summarized in the final.

\subsection{Discussion on $\mathbf u^1=\mathbf w^1$ satisfying the boundary condition Eq.~(\ref{Boundary1})\label{goddamn1}}

\paragraph*{$l_4\neq 0$} For the boundary base $\mathbf w^1$ with entries
\begin{gather}
	\mathbf w^1=\begin{pmatrix}
         \beta_1w^{1,a} & \beta_1w^{1,b} & \beta_1^2 w^{1,a} & \beta_1^2 w^{1,b} & \cdots & \beta_1^{N} w^{1,a} & \beta_1^{N} w^{1,b}
    \end{pmatrix}^\mT,\label{Cminor}
\end{gather}
the bulk recurrence relation Eq.~(\ref{Bulk1}) is solved by
\begin{gather}
    w^{1,a}=l_4(1+\beta_1^2)c_1,\quad 
    w^{1,b}=-(l_3+m\beta_1+l_2\beta_1^2) c_1,\notag
\end{gather}
where $c_1\neq 0$ is a normalization constant. Since $0<|\beta_1|<1$, $\mathbf w^1$ is required to satisfy the boundary condition Eq.~(\ref{Boundary1}), which renders
\begin{subequations}
	\begin{gather}
\beta_1(\beta _1^3 \left(l_1 l_4-l_2 l_3\right)+\beta _1 \left(-l_3^2+l_1 l_4-m^2\right)+\beta _1^2 \left(-ml_2-l_3 m+l_4 n\right)-l_3 m+l_4 n)=0,
\label{webern}\\
 \beta_1l_4 (\beta_1(l_2-l_3)+m)=0. \label{mahler}
	\end{gather}
\end{subequations}
Eq.~(\ref{mahler}) implies $\beta_1(l_2-l_3)+m=0$, rendering either $\left\{\begin{array}{l}
	l_2 = l_3\\ m = 0
\end{array}\right.$  or  $\beta_1 = \frac{m}{l_3 - l_2}$.
$\left\{\begin{array}{l}
	l_2 = l_3\\ m = 0
\end{array}\right.$ together with Eq.~(\ref{webern})  enforces $|\beta_1| = 1$ or $|\beta_1|=0$, which does not correspond to a boundary zero mode and should be discarded. 
The physical meaning of $m$ also forbids $m=0$. 
Hence we shall continue with $\beta_1 = \frac{m}{l_3 - l_2}$. Since $l_4\neq 0$, we can safely set $l_1=l_2l_3/l_4$. Eq.~(\ref{webern}) 
and Eq.~(\ref{damn1}) 
render
\begin{gather}
     \frac{\left((l_2-l_3)^2+m^2\right) (2 l_3 m-l_4 n)}{(l_2-l_3)^2}=0.
     \notag
\end{gather}
Therefore, if $2l_3m - l_4n = 0$, Eq.~(\ref{Cminor}) represents a boundary state that satisfies the boundary condition Eq.~(\ref{Boundary1}). Taking the aforementioned conditions into account, we obtain that if
\begin{gather}
    \mathbb A_1:~\left\{\begin{array}{l}
    m\neq 0,\quad l_4\neq 0,\quad l_2\neq l_3,\\
    l_1l_4=l_2l_3,\quad
    2l_3m=l_4n,\\
    |m|<|l_3-l_2|,
    \end{array}\right. 
\end{gather}
then there exists one boundary zero mode $\mathbf u^1=\mathbf w^1$ given by Eq.~(\ref{Cminor}), where
\begin{gather}
    w^{1,a}=\frac{l_4}{\beta_1\sqrt{l_3^2+l_4^2}}\sqrt{\frac{1-\beta_1^2}{1-\beta_1^{2N}}} ,\quad 
    w^{1,b}=-\frac{l_3}{\beta_1\sqrt{l_3^2+l_4^2}}\sqrt{\frac{1-\beta_1^2}{1-\beta_1^{2N}}},\quad \beta_1 = \frac{m}{l_3 - l_2} .\notag
\end{gather}
\paragraph*{$l_4=0$} The above discussion ignored $l_4=0$. If $l_4=0$, $\det B=0$ necessitates $l_2l_3=0$. 

If $l_2=l_3=l_4=0$, i.e.
\begin{gather}
    A=\begin{pmatrix}
        n& m \\ m& 0
    \end{pmatrix},\quad B=\begin{pmatrix}
        l_1 & 0 \\ 0 & 0
    \end{pmatrix},\notag
\end{gather}
the bulk recurrence relation Eq.~(\ref{Bulk1}) and the boundary condition Eq.~(\ref{Boundary1}) render
\begin{gather}
	\begin{pmatrix}
		l_1+n\beta_1+l_1\beta_1^2 & m\beta_1 \\ m\beta_1 & 0
	\end{pmatrix}\begin{pmatrix}
		w^{1,a} \\ w^{1,b}
	\end{pmatrix}=0,\quad \begin{pmatrix}
		\beta_1(n+l_1\beta_1) & m\beta_1 \\ m\beta_1 & 0
	\end{pmatrix}\begin{pmatrix}
		w^{1,a} \\ w^{1,b}
	\end{pmatrix}=0.\label{Cmajor}
\end{gather}
Nontrivial solution to Eq.~(\ref{Cmajor}) enforces $m=0$, resulting in a zero determinant for the $K$ matrix. Therefore, $l_2=l_3=l_4=0$ is excluded from consideration.

If $l_2=0$, $l_3\neq 0$, i.e.
\begin{gather}
    A=\begin{pmatrix}
        n& m \\ m& 0
    \end{pmatrix},\quad B=\begin{pmatrix}
        l_1 & 0 \\ l_3 & 0
    \end{pmatrix},\notag
\end{gather}
for $\mathbf w^1$, the bulk recurrence relation Eq.~(\ref{Bulk1}) is solved by
\begin{gather}
    w^{1,a}=0,\quad w^{1,b}=c_1,\quad \beta_1=-\frac{m}{l_3} .\label{siu}
\end{gather}
where $c_1\neq 0$ is the normalization constant. Since $0<|\beta_1|<1$, $\mathbf w^1$ is required to satisfy the boundary condition Eq.~(\ref{Boundary1}), which gives
\begin{gather}
    \begin{pmatrix}
        \beta_1 ^2 l_1+\beta_1 n & \beta_1 ^2 l_3+\beta_1  m \\
        \beta_1 m & 0
    \end{pmatrix}\begin{pmatrix}
        w^{1,a} \\ w^{1,b}
    \end{pmatrix}=0.\label{siuu}
\end{gather}
If $\mathbf w^1$ is solved by Eq.~(\ref{siu}), Eq.~(\ref{siuu}) naturally holds. Therefore, if
\begin{gather}
    \mathbb A_2:~\left\{\begin{array}{l}
    m\neq 0,\quad
    l_3\neq 0,\\
    l_4=l_2= 0,\\
         |m|<|l_3|,
    \end{array}\right.
\end{gather}
there exists a boundary zero mode $\mathbf u^1=\mathbf w^1$ given by Eq.~(\ref{Cminor}), satisfying the boundary condition Eq.~(\ref{Boundary1}), where
\begin{gather}
    w^{1,a}=0,\quad w^{1,b}=\frac{1}{\beta_1}\sqrt{\frac{1-\beta_1^{2}}{1-\beta_1^{2N}}},\quad \beta_1=-\frac{m}{l_3}.\notag
\end{gather}

If $l_3=0$, $l_2\neq 0$, i.e.
\begin{gather}
    A=\begin{pmatrix}
        n& m \\ m& 0
    \end{pmatrix},\quad B=\begin{pmatrix}
        l_1 & l_2 \\ 0 & 0
    \end{pmatrix},\notag
\end{gather}
the bulk recurrence relation Eq.~(\ref{Bulk1}) is solved by
\begin{gather}
    w^{1,a}=-(l_2+m\beta_1)c_1,\quad 
    w^{1,b}=(l_1+n\beta_1+l_1\beta_1^2)c_1,\quad 
    \beta_1=-\frac{m}{l_2},\notag
\end{gather}
where $c_1\neq 0$ is a normalization constant. Since $0<|\beta_1|<1$, $\mathbf w^1$ is required to satisfy the boundary condition Eq.~(\ref{Boundary1}), which renders
\begin{gather}
	\begin{pmatrix}
		n\beta_1+l_1\beta_1^2 & m\beta_1 \\ m\beta_1+l_2\beta_1^2 & 0
	\end{pmatrix}\begin{pmatrix}
        w^{1,a} \\ w^{1,b}
    \end{pmatrix}=0.\notag
\end{gather}
Therefore, if the elements of the $K$ matrix in Eq.~(\ref{ktridiagonal}) satisfy
\begin{gather}
	\mathbb  A_3:~\left\{\begin{array}{l}
    m\neq 0,\quad l_2\neq 0,\\
    nl_2=2l_1m,\\
    l_3=l_4=0,\\
    |m|<|l_2|,
    \end{array}\right.
\end{gather}
then a boundary zero mode $\mathbf u^1=\mathbf w^1$ given by Eq.~(\ref{Cminor}) exists that match the boundary condition Eq.~(\ref{Boundary1}), where
\begin{gather*}
	w^{1,a}=\frac{2 \left(m^2-l_2^2\right)}{\beta_1\sqrt{(m n-2 l_1 l_2)^2+4 \left(l_2^2-m^2\right)^2}}\sqrt{\frac{1-\beta_1^{2}}{1-\beta_1^{2N}}},\  w^{1,b}=\frac{2 l_1 l_2-m n}{\beta_1\sqrt{(m n-2 l_1 l_2)^2+4 \left(l_2^2-m^2\right)^2}}\sqrt{\frac{1-\beta_1^{2}}{1-\beta_1^{2N}}},\ \beta_1=-\frac{m}{l_2}.\\
\end{gather*}

\ 

\subsection{Discussion on $\mathbf u^2=\mathbf w^2$ satisfying the boundary condition Eq.~(\ref{Boundary2})\label{goddamn2}}
\paragraph*{$l_4\neq 0$} 
Likewise, if $l_4\neq 0$, for the boundary base $\mathbf w^2$ with entries
\begin{gather}
	\mathbf w^2=\begin{pmatrix}
         \beta_2w^{2,a} & \beta_2w^{2,b} & \beta_2^2 w^{2,a} & \beta_2^2 w^{2,b} & \cdots & \beta_2^{N} w^{2,a} & \beta_2^{N} w^{2,b}
    \end{pmatrix}^\mT,\label{Dmajor}
\end{gather}
the bulk recurrence relation Eq.~(\ref{Bulk1}) is solved by
\begin{gather}
    w^{2,a}=l_4(1+\beta_2^2)c_2 ,\quad
    w^{2,b}=-(l_3+m\beta_2+l_2\beta_2^2) c_2, \notag
\end{gather}
where $c_2\neq 0$ is a normalization constant. Since $|\beta_2|>1$, it is required to satisfy the boundary condition Eq.~(\ref{Boundary2}), rendering
\begin{subequations}
\begin{gather}
\beta _2^3 \left(l_2 m-l_4 n+m^2\right)+\beta _2^2 \left(l_3 m+l_2^2-l_1 l_4\right)+\beta _2 \left(l_2 m-l_4 n\right)+l_2 l_3-l_1 l_4=0,
\label{slithering}\\
 l_4 \beta_2^2 ((l_3-l_2)+m\beta_2) =0.
    \label{bach}
\end{gather}
\end{subequations}
Eq.~(\ref{bach}) implies $(l_3-l_2)+m\beta_2=0$, which enforces either $\left\{\begin{array}{l}
	l_2 = l_3\\ m = 0
\end{array}\right.$  or  $\beta_2 = \frac{l_2-l_3}{m}$.
$\left\{\begin{array}{l}
	l_2 = l_3\\ m = 0
\end{array}\right.$   enforces $|\beta_2|=0$ or $|\beta_2| = 1$, which does not correspond to a boundary zero mode and should be discarded. Hence we shall continue with $\beta_2=\frac{l_2-l_3}{m}$. Since $l_4\neq 0$, we can safely set $l_1=l_2l_3/l_4$. Eq.~(\ref{slithering}) 
and Eq.~(\ref{damn1}) 
render
\begin{gather}
    \frac{\left((l_2-l_3)^2+m^2\right) (2 l_2 m-l_4 n)}{(l_2-l_3)^2}=0.\notag 
 \end{gather}
Therefore, if $2l_2m-l_4n=0$, Eq.~(\ref{Dmajor}) represents a boundary zero mode that satisfies the boundary condition Eq.~(\ref{Boundary2}). Taking the aforementioned conditions into account, if
\begin{gather}
    \mathbb B_1:~\left\{\begin{array}{l}
    m\neq 0,\quad 
    l_4\neq 0,\\
    l_1l_4=l_2l_3,\quad
    2l_2m=l_4n,\\
    |m|<|l_3-l_2|,
    \end{array}\right. 
\end{gather}
there exists one boundary zero mode $\mathbf u^2=\mathbf w^2$ given by Eq.~(\ref{Dmajor}), where
\begin{gather}
    w^{2,a}=\frac{l_4}{\beta_2\sqrt{l_2^2+l_4^2}}\sqrt{\frac{1-\beta_2^2}{1-\beta_2^{2N}}},\quad 
    w^{2,b}=-\frac{l_2}{\beta_2\sqrt{l_2^2+l_4^2}}\sqrt{\frac{1-\beta_2^2}{1-\beta_2^{2N}}},\quad \beta_2=\frac{l_2-l_3}{m}. \notag
\end{gather}

\paragraph*{$l_4=0$} 
If $l_4=0$, $\det B=0$ enforces $l_2l_3=0$.

If $l_2=l_3=l_4=0$, i.e.
\begin{gather}
    A=\begin{pmatrix}
        n& m \\ m& 0
    \end{pmatrix},\quad B=\begin{pmatrix}
        l_1 & 0 \\ 0 & 0
    \end{pmatrix},\notag
\end{gather}
the bulk recurrence relation Eq.~(\ref{Bulk1}) and the boundary condition Eq.~(\ref{Boundary2}) render
\begin{gather}
	\begin{pmatrix}
		l_1+n\beta_1+l_1\beta_1^2 & m\beta_1 \\ m\beta_1 & 0
	\end{pmatrix}\begin{pmatrix}
		w^{1,a} \\ w^{1,b}
	\end{pmatrix}=0,\quad \begin{pmatrix}
		l_1+n\beta_1 & m\beta_1 \\ m\beta_1 & 0
	\end{pmatrix}\begin{pmatrix}
		w^{1,a} \\ w^{1,b}
	\end{pmatrix}=0.\label{Dminor}
\end{gather}
Nontrivial solution to Eq.~(\ref{Dminor}) enforces $m=0$, resulting in a zero determinant for the $K$ matrix. Therefore, $l_2=l_3=l_4=0$ is excluded from consideration.

If $l_3=0$, $l_2\neq 0$, i.e.
\begin{gather}
    A=\begin{pmatrix}
        n& m \\ m& 0
    \end{pmatrix},\quad B=\begin{pmatrix}
        l_1 & l_2 \\ 0 & 0
    \end{pmatrix},\notag
\end{gather}
for $\mathbf w^2$, the bulk recurrence relation Eq.~(\ref{Bulk1}) is solved by
\begin{gather}
    w^{2,a}=0,\quad w^{2,b}=c_2,\quad \beta_2=-\frac{l_2}{m},\label{siu2}
\end{gather}
where $c_2\neq 0$ is a normalization constant. Since $|\beta_2|>1$, $\mathbf w^2$ is required to satisfy the boundary condition Eq.~(\ref{Boundary2}), which gives
\begin{gather}
    \begin{pmatrix}
        l_1+\beta_2 n & l_2+\beta_2 m\\
        \beta_2 m & 0
    \end{pmatrix}\begin{pmatrix}
        w^{2,a}\\ w^{2,b}
    \end{pmatrix}=0. \label{siuu2}
\end{gather}
If $\mathbf w^2$ is solved by Eq.~(\ref{siu2}), Eq.~(\ref{siuu2}) naturally holds. Therefore, if
\begin{gather}
    \mathbb B_2:~\left\{\begin{array}{l}
    m\neq 0,\quad 
    l_2\neq 0,\\
    l_4=l_3= 0,\\
         |m|<|l_2|,
    \end{array}\right. 
\end{gather}
there exists a boundary zero mode $\mathbf u^2=\mathbf w^2$ with
\begin{gather}
    w^{2,a}=0,\quad w^{2,b}=\frac{1}{\beta_2}\sqrt{\frac{1-\beta_2^{2}}{1-\beta_2^{2N}}},\quad \beta_2=-\frac{l_2}{m},\notag
\end{gather}
satisfying the boundary condition Eq.~(\ref{Boundary2}).

If $l_2=0$, $l_3\neq 0$, i.e.
\begin{gather}
    A=\begin{pmatrix}
        n& m \\ m& 0
    \end{pmatrix},\quad B=\begin{pmatrix}
        l_1 & 0 \\ l_3 & 0
    \end{pmatrix},\notag
\end{gather}
the bulk recurrence relation Eq.~(\ref{Bulk1}) is solved by
\begin{gather}
    w^{2,a}=-(m+l_3\beta_2) c_2,\quad 
    w^{2,b}=(l_1+n\beta_2+l_1 \beta_2^2)c_2,\quad
    \beta_2=-\frac{l_3}{m},\notag
\end{gather}
where $c_2\neq 0$ is a normalization constant. 
Since $|\beta_2|>1$, $\mathbf w^2$ is required to satisfy the boundary condition specified by Eq.~(\ref{Boundary2}), which renders
\begin{gather}
	\left(
\begin{array}{cc}
 l_1+\beta_2  n & l_3+\beta_2  m \\
 \beta_2  m & 0 \\
\end{array}
\right)\begin{pmatrix}
	w^{2,a}\\ w^{2,b}
\end{pmatrix}=0.
\end{gather}
Therefore, if the elements of the $K$ matrix in Eq.~(\ref{ktridiagonal}) satisfy
\begin{gather}
	\mathbb B_3:~\left\{\begin{array}{l}
    m\neq 0,\quad l_3\neq 0,\\
    nl_3=2l_1m,\\
    l_2=l_4=0,\\
    |m|<|l_3|,
    \end{array}\right.
\end{gather}
then a boundary zero mode $\mathbf u^2=\mathbf w^2$ given by Eq.~(\ref{Cminor}) exists that match the boundary condition Eq.~(\ref{Boundary2}), where
\begin{gather}
	w^{2,a}=\frac{l_1 l_3-\frac{m n}{2}}{\beta_2\sqrt{\left(l_1 l_3-\frac{m n}{2}\right)^2+\left(l_3^2-m^2\right)^2}}\sqrt{\frac{1-\beta_2^{2}}{1-\beta_2^{2N}}},\ w^{2,b}=\frac{l_3^2-m^2}{\beta_2\sqrt{\left(l_1 l_3-\frac{m n}{2}\right)^2+\left(l_3^2-m^2\right)^2}}\sqrt{\frac{1-\beta_2^{2}}{1-\beta_2^{2N}}},\ \beta_2=-\frac{l_3}{m}.
\end{gather}

\subsection{The conditions for boundary zero modes under $\det B=0$\label{goddamn3}}
The set operation $\mathbb A_1\cup \mathbb A_2 \cup \mathbb A_3$ yields the conditions for the appearance of a boundary zero mode that satisfies the boundary condition Eq.~(\ref{Boundary1}):
\begin{gather}
	\mathbb A_1:~\left\{\begin{array}{l}
    m\neq 0,\quad l_4\neq 0,\quad l_2\neq l_3,\\
    l_1l_4=l_2l_3,\quad
    2l_3m=l_4n,\\
    |m|<|l_3-l_2|,
    \end{array}\right. \quad \text{or}\quad \mathbb A_2:~\left\{\begin{array}{l}
    m\neq 0,\quad
    l_3\neq 0,\\
    l_4=l_2= 0,\\
         |m|<|l_3|,
    \end{array}\right. \quad \text{or} \quad  \mathbb A_3:~\left\{\begin{array}{l}
    m\neq 0,\quad l_2\neq 0,\\
    nl_2=2l_1m,\quad
    l_3=l_4=0,\\
    |m|<|l_2|.
    \end{array}\right.
\end{gather}
Likewise, the set operation $\mathbb B_1 \cup \mathbb B_2 \cup \mathbb B_3$ yields the conditions for the appearance of a boundary zero mode that satisfies the boundary condition Eq.~(\ref{Boundary2}):
\begin{gather}
	\mathbb B_1:~\left\{\begin{array}{l}
    m\neq 0,\quad 
    l_4\neq 0,\quad l_2\neq l_3,\\
    l_1l_4=l_2l_3,\quad
    2l_2m=l_4n,\\
    |m|<|l_3-l_2|,
    \end{array}\right. 
    \quad \text{or} \quad
    \mathbb B_2:~\left\{\begin{array}{l}
    m\neq 0,\quad 
    l_2\neq 0,\\
    l_4=l_3= 0,\\
         |m|<|l_2|,
    \end{array}\right. 
    \quad \text{or} \quad
   \mathbb B_3:~ \left\{\begin{array}{l}
    m\neq 0,\quad l_3\neq 0,\\
    nl_3=2l_1m,\quad
    l_2=l_4=0,\\
    |m|<|l_3|.
    \end{array}\right.
\end{gather}
Taking all the conditions for boundary zero modes into account, we conclude that the 
$K$ matrix in Eq.~(\ref{ktridiagonal}) has one boundary zero mode if one of the following conditions is satisfied:
\begin{gather}
    \left\{\begin{array}{l}
    m\neq 0,\quad l_4\neq 0,\\
    l_1l_4=l_2l_3,\\
    2l_3m=l_4n\ \text{ or }\ 2l_2m=l_4n,\\
    |m|<|l_3-l_2|,
    \end{array}\right.\text{ or }\left\{\begin{array}{l}
    m\neq 0,\quad
    l_3\neq 0,\\
    l_4=l_2= 0,\\
    nl_3 \neq 2l_1 m\\
         |m|<|l_3|,
    \end{array}\right.\text{ or }\left\{\begin{array}{l}
    m\neq 0,\quad
    l_2\neq 0,\\
    l_4=l_3= 0,\\
    nl_2 \neq 2l_1 m\\
         |m|<|l_2|,
    \end{array}\right. \label{janacek1}
\end{gather}
which is obtained by the set operation $\mathbb V \backslash [(\mathbb A_1 \cup \mathbb A_2 \cup\mathbb A_3)\cap(\mathbb B_1 \cup \mathbb B_2 \cup \mathbb B_3)]$, where $\mathbb V=\mathbb A_1 \cup \mathbb A_2 \cup\mathbb A_3 \cup \mathbb B_1 \cup \mathbb B_2 \cup\mathbb B_3$.
More specifically, if $\left\{\begin{array}{l}l_4=0\\|l_2|<|l_3|\end{array}\right.\text{ or }\left\{\begin{array}{l}l_4\neq 0\\2l_3m=l_4n\end{array}\right.$, the boundary zero mode is a lower boundary zero mode, satisfying the boundary condition Eq.~(\ref{Boundary1}). If $\left\{\begin{array}{l}l_4=0\\|l_3|<|l_2|\end{array}\right.\text{ or }\left\{\begin{array}{l}l_4\neq 0\\2l_2m=l_4n\end{array}\right.$, the boundary zero mode is an upper boundary zero mode, satisfying the boundary condition Eq.~(\ref{Boundary2}).

Moreover, if the elements of the $K$ matrix satisfies
\begin{gather}
    \left\{\begin{array}{l}
    m\neq 0,\quad l_2\neq 0,\\
    l_3=l_4=0,\\
    nl_2=2l_1m,\\
    |m|<|l_2|,
    \end{array}\right.\text{ or }\left\{\begin{array}{l}
    m\neq 0,\quad l_3\neq 0,\\
    l_2=l_4=0,\\
    nl_3=2l_1m,\\
    |m|<|l_3|,
    \end{array}\right.\label{janacek2}
\end{gather}
the $K$ matrix has two boundary zero modes, which is obtained by the set operation $(\mathbb A_1\cup\mathbb  A_2\cup \mathbb A_3)\cap (\mathbb B_1\cup\mathbb  B_2\cup \mathbb B_3)$.

\section{The Conditions for the Existence of Boundary Zero Modes under $\det B\neq 0$\label{section2}}
The parameters $\{\beta_j\}$ in the boundary base (Eq.~(\ref{phij})) controls the boundary condition that a trial solution (Eq.~(\ref{lifesucks1})) is required to satisfy. 
In the discussion of $\det B\neq 0$ cases, the solutions for $\beta_j$ ($j=1,2,3, 4$) are provided by Eq.~(\ref{equationforbeta11}), which is a quartic equation and the solutions for $\beta$ are given by:
\begin{gather*}
    \beta_1=\frac{-\alpha_1 \sqrt{\frac{2 \alpha_2 \left(\sqrt{8 \alpha_1^2-4 \alpha_1 \alpha_3+\alpha_2^2}+\alpha_2\right)-4\alpha_1 \alpha_3}{\alpha_1^2}-8}-\sqrt{8 \alpha_1^2-4 \alpha_1 \alpha_3+\alpha_2^2}-\alpha_2}{4\alpha_1},\\
    \beta_2=\frac{\alpha_1 \sqrt{\frac{2 \alpha_2 \left(\sqrt{8 \alpha_1^2-4 \alpha_1 \alpha_3+\alpha_2^2}+\alpha_2\right)-4\alpha_1 \alpha_3}{\alpha_1^2}-8}-\sqrt{8 \alpha_1^2-4 \alpha_1 \alpha_3+\alpha_2^2}-\alpha_2}{4\alpha_1},\\
    \beta_3=\frac{-\alpha_1 \sqrt{\frac{2 \alpha_2 \left(\alpha_2-\sqrt{8 \alpha_1^2-4 \alpha_1 \alpha_3+\alpha_2^2}\right)-4\alpha_1 \alpha_3}{\alpha_1^2}-8}+\sqrt{8 \alpha_1^2-4 \alpha_1 \alpha_3+\alpha_2^2}-\alpha_2}{4\alpha_1},\\
    \beta_4=\frac{\alpha_1 \sqrt{\frac{2 \alpha_2 \left(\alpha_2-\sqrt{8 \alpha_1^2-4 \alpha_1 \alpha_3+\alpha_2^2}\right)-4\alpha_1 \alpha_3}{\alpha_1^2}-8}+\sqrt{8 \alpha_1^2-4 \alpha_1 \alpha_3+\alpha_2^2}-\alpha_2}{4\alpha_1}.
\end{gather*}
These solutions render the relation
\begin{gather}
    \beta_1\beta_2=1,\quad \beta_3\beta_4=1.\label{betarelation}
\end{gather}
If $l_4\neq 0$, carrying on a comprehensive analytical analysis is challenging and requires a case-by-case study, which is left for future research. To simplify discussion, we focus on the special case where $l_4=0$.
Setting $l_4=0$, Eq.~(\ref{equationforbeta11}) becomes
\begin{gather}
    (l_2+m\beta+l_3\beta^2)(l_3+m\beta+l_2\beta^2)=0.\label{l4=0betasolution}
\end{gather}
$\det B\neq 0$ together with $l_4=0$ renders $l_2l_3\neq 0$. Hence the solutions to Eq.~(\ref{l4=0betasolution}) are
\begin{subequations}
	\begin{gather}
    \beta_1=\frac{-m-\sqrt{m^2-4l_2l_3}}{2l_3},\label{beta1}\\
    \beta_2=\frac{-m+\sqrt{m^2-4l_2l_3}}{2l_2},\label{beta2}\\
    \beta_3=\frac{-m+\sqrt{m^2-4l_2l_3}}{2l_3},\label{beta3}\\
    \beta_4=\frac{-m-\sqrt{m^2-4l_2l_3}}{2l_2}.\label{beta4}
\end{gather}
\end{subequations}
$\beta_1$ and $\beta_3$ are solutions to $l_2+m\beta+l_3\beta^2=0$. $\beta_2$ and $\beta_4$ are solutions to $l_3+m\beta+l_2\beta^2=0$. Moreover, $\beta_1\beta_2=1$ and $\beta_3\beta_4=1$ hold.
It can be easily verified that $|\beta_1|=|\beta_2|=1$ enforces $|\beta_3|=|\beta_4|=1$, and vice versa. Therefore, we only need to consider 4 cases:

Case I. $|\beta_1|>1$, $|\beta_2|<1$, $|\beta_3|>1$, $|\beta_4|<1$;

Case II. $|\beta_1|<1$, $|\beta_2|>1$, $|\beta_3|<1$, $|\beta_4|>1$;

Case III. $|\beta_1|>1$, $|\beta_2|<1$, $|\beta_3|<1$, $|\beta_4|>1$;

Case IV. $|\beta_1|<1$, $|\beta_2|>1$, $|\beta_3|>1$, $|\beta_4|<1$.

In the subsequent analysis, we divide our discussion into two subsections addressing boundary conditions Eq.~(\ref{Boundary1}) and Eq.~(\ref{Boundary2}), respectively.
Within each subsection, the discussion is organized into four distinct cases: Case I $\sim$ Case IV. 
For each case, trial solutions are constructed using the boundary bases. 
Each trial solution, denoted as $\mathbf u^{i,j}$, is a linear superposition of two boundary bases, $\mathbf w^i$ and $\mathbf w^j$ (Eq.~(\ref{phij})), where the moduli of parameters $\beta_i$ and $\beta_j$ are either both greater than 1 or both less than 1. These trial solutions are required to match the corresponding boundary conditions.
This procedure yields explicit constraints on the elements of the $K$ matrix that guarantee the existence of boundary zero modes.

\subsection{Discussion on the boundary condition Eq.~(\ref{Boundary1})\label{chewinggum1}}

\paragraph*{Case I} $|\beta_1|>1$, $|\beta_2|<1$, $|\beta_3|>1$, $|\beta_4|<1$.

Since $|\beta_2|<1$, $|\beta_4|<1$, $\mathbf{u}^{2,4}$ is required to satisfy the boundary condition Eq.~(\ref{Boundary1}).
The boundary bases corresponding to $\beta_2$ and $\beta_4$ are denoted as $\mathbf w^2$ and $\mathbf w^4$, respectively.
\begin{gather*}
    \mathbf{w}^j=\begin{pmatrix}
        \beta_j w^{j,a} & \beta_j w^{j,b} & \beta_j^2 w^{j,a} & \beta_j^2 w^{j,b} & \cdots & \beta_j^N w^{j,a} &\beta_j^N w^{j,b}
    \end{pmatrix}^\mT,\quad j=2,4.
\end{gather*}
The bulk recurrence relation Eq.~(\ref{Bulk1}) gives
\begin{gather}
    \begin{pmatrix}
        l_1+n\beta_{j}+l_1\beta_{j}^2 & l_2+m\beta_{j}+l_3\beta_{j}^2\\
        0 & 0
    \end{pmatrix}\begin{pmatrix}
        w^{j,a}\\ w^{j,b}
    \end{pmatrix}=0,\quad j=2,4.\label{phi2phi4bulk1}
\end{gather}
Eq.~(\ref{phi2phi4bulk1}) is solved by
\begin{gather*}
    w^{j,a}=l_2+m\beta_j+l_3\beta_j^2,\quad 
    w^{j,b}=-(l_1+n\beta_j+l_1\beta_j^2),\quad j=2,4.
\end{gather*}
$\mathbf{u}^{2,4}$ is a linear combination of $\mathbf{w}^2$ and $\mathbf{w}^4$.
\begin{gather*}
    \mathbf{u}^{2,4}=c_2\mathbf w^2+c_4\mathbf w^4=\begin{pmatrix}
        u^{2,4,A}_1 & u^{2,4,B}_1 & u^{2,4,A}_2 &  u^{2,4,B}_2 & \cdots & u^{2,4,A}_N &  u^{2,4,B}_N
    \end{pmatrix}^\mT,
\end{gather*}
where
\begin{gather*}
    \begin{pmatrix}
        u_l^{2,4,a}\\
        u_l^{2,4,b}
    \end{pmatrix}=\beta_2^l\begin{pmatrix}
        l_2+m\beta_2+l_3\beta_2^2 \\ -(l_1+n\beta_2+l_1\beta_2^2)
    \end{pmatrix}c_2+\beta_4^l\begin{pmatrix}
        l_2+m\beta_4+l_3\beta_4^2 \\ -(l_1+n\beta_4+l_1\beta_4^2)
    \end{pmatrix}c_4.
\end{gather*}
The boundary condition Eq.~(\ref{Boundary1}) renders
\begin{gather*}
    \left(
\begin{array}{cc}
 \beta _2^2 \left(l_1 l_2-l_1 l_3\right)+\beta _2 \left(l_2 n-l_1 m\right) & \beta _4^2 \left(l_1 l_2-l_1 l_3\right)+\beta _4 \left(l_2 n-l_1 m\right) \\
 \beta_2^2 l_2 \left(l_2+\beta_2^2 l_3+\beta_2 m\right)+\beta_2 m \left(l_2+\beta_2^2 l_3+\beta_2 m\right) & \beta_4^2 l_2 \left(l_2+\beta_4^2 l_3+\beta_4 m\right)+\beta_4 m \left(l_2+\beta_4^2 l_3+\beta_4 m\right) \\
\end{array}
\right)\begin{pmatrix}
    c_2 \\ c_4
\end{pmatrix}=0.
\end{gather*}
Nontrivial solution to $c_2$, $c_4$ renders
\begin{gather*}
    (l_2-l_3)(n(l_2+l_3)-2l_1 m)=0.
\end{gather*}
$l_2=l_3$ renders either $|\beta_2||\beta_4|=1$ or $|\beta_2|=|\beta_4|=1$, contradicting the assumption that $|\beta_2|<1,|\beta_4|<1$. Therefore, if the elements of the $K$  matrix satisfy
\begin{gather}
\mathbb C_1:~ \left\{
\begin{array}{l}
l_4=0,\\
m\neq 0,\quad  l_2\neq l_3,\quad l_2l_3\neq 0 \\
|\beta_1|>1, |\beta_2|<1, |\beta_3|>1, |\beta_4|<1,\\
	n(l_2+l_3)=2l_1 m, 	
\end{array}
\right.\label{s1}
\end{gather}
there exists a boundary zero mode satisfying boundary condition Eq.~(\ref{Boundary1}).
 
 %
  
  \
 
 \paragraph*{Case II} $|\beta_1|<1$, $|\beta_2|>1$, $|\beta_3|<1$, $|\beta_4|>1$.
 
Since $|\beta_1|<1$, $|\beta_3|<1$, $\mathbf{u}^{1,3}$ is required to satisfy the boundary condition Eq.~(\ref{Boundary1}).
Denote the boundary bases corresponding to $\beta_1$ and $\beta_3$ as $\mathbf w^1$ and $\mathbf w^3$, respectively.
\begin{gather*}
    \mathbf{w}^j=\begin{pmatrix}
        \beta_j w^{j,a} & \beta_j w^{j,b} & \beta_j^2 w^{j,a} & \beta_j^2 w^{j,b} & \cdots & \beta_j^N w^{j,a} &\beta_j^N w^{j,b}
    \end{pmatrix}^\mT,\quad j=1,3.
\end{gather*}
The bulk recurrence relation Eq.~(\ref{Bulk1}) gives
\begin{gather}
     \begin{pmatrix}
        l_1+n\beta_j+l_1\beta_j^2 & 0 \\
        l_3+m\beta_j+l_2\beta_j^2 & 0 
    \end{pmatrix}\begin{pmatrix}
        w^{j,a}\\ w^{j,b}
    \end{pmatrix}=0,\quad j=1,3.\label{phi1phi3bulk2}
\end{gather}
Eq.~(\ref{phi1phi3bulk2}) is solved by
\begin{gather}
	w^{j,a}=0,\quad w^{j,b}=1,\quad j=1,3.
\end{gather}
$\mathbf u^{1,3}$ is a linear combination of $\mathbf{w}^1$ and $\mathbf{w}^3$.
\begin{gather*}
    \mathbf{u}^{1,3}=c_1\mathbf w^1+c_3\mathbf w^3=\begin{pmatrix}
        u^{1,3,a}_1 & u^{1,3,b}_1 & u^{1,3,a}_2 &u^{1,3,b}_2 &\cdots & u^{1,3,a}_N & u^{1,3,b}_N
    \end{pmatrix}^\mT,
\end{gather*}
where
\begin{gather}
	\begin{pmatrix}
        u_l^{1,3,a}\\
        u_l^{1,3,b}
    \end{pmatrix}=\beta_1^l\begin{pmatrix}
        0\\
        1
    \end{pmatrix}c_1+\beta_3^l\begin{pmatrix}
        0\\
        1
    \end{pmatrix}c_3.
\end{gather}
The boundary condition Eq.~(\ref{Boundary1}) renders
\begin{gather}
    \left(
\begin{array}{cc}
 \beta_1 (\beta_1 l_3+m) & \beta_3 (\beta_3 l_3+m) \\
 0 & 0 \\
\end{array}
\right)\begin{pmatrix}
    c_1 \\ c_3
\end{pmatrix}=0. \label{fledge}
\end{gather}
Eq.~(\ref{fledge}) always has nontrivial solutions to $c_1$ and $c_3$.
\begin{gather}
	c_1=-\beta_3(\beta_3 l_3+m)c,\quad c_3=\beta_1(\beta_1 l_3+m)c,
\end{gather}
where $c$ is a normalization constant.
Therefore, there exists a boundary zero mode satisfying the boundary condition Eq.~(\ref{Boundary1}) under the condition
\begin{gather}
	\mathbb C_2: ~ \left\{\begin{array}{l}
	    l_4=0,\\
		m\neq 0,\quad  l_2l_3\neq 0,\\
		|\beta_1|<1, |\beta_2|>1, |\beta_3|<1, |\beta_4|>1.
	\end{array}\right.\label{s2}
\end{gather}

\

\paragraph*{Case III} $|\beta_1|>1$, $|\beta_2|<1$, $|\beta_3|<1$, $|\beta_4|>1$.

Since $|\beta_2|<1,|\beta_3|<1$, $\mathbf u^{2,3}$ is required to satisfy the boundary condition specified by Eq.~(\ref{Boundary1}).
The boundary bases corresponding to $\beta_2$ and $\beta_3$ are denoted as $\mathbf w^2$ and $\mathbf w^3$, respectively.
\begin{gather*}
    \mathbf{w}^j=\begin{pmatrix}
        \beta_j w^{j,a} & \beta_j w^{j,b} & \beta_j^2 w^{j,a} & \beta_j^2 w^{j,b} & \cdots & \beta_j^N w^{j,a} &\beta_j^N w^{j,b}
    \end{pmatrix}^\mT,\quad j=2,3.
\end{gather*}
The bulk recurrence relation Eq.~(\ref{Bulk1}) gives
\begin{gather}
    \begin{pmatrix}
        l_1+n\beta_2+l_1\beta_2^2 & l_2+m\beta_2+l_3\beta_2^2 \\
        0 & 0 
    \end{pmatrix}\begin{pmatrix}
        w^{2,a}\\ w^{2,b}
    \end{pmatrix}=0,\quad 
    \begin{pmatrix}
        l_1+n\beta_3+l_1\beta_3^2 & 0 \\
        l_3+m\beta_3+l_2\beta_3^2 & 0 
    \end{pmatrix}\begin{pmatrix}
        w^{3,a}\\ w^{3,b}
    \end{pmatrix}=0. \label{phi2phi3bulk1}
\end{gather}
The solution to Eq.~(\ref{phi2phi3bulk1}) is given by
\begin{gather*}
\left\{\begin{array}{l}
     w^{2,a}=l_2+m\beta_2+l_3\beta_2^2 \\
     w^{2,b}=-(l_1+n\beta_2+l_1\beta_2^2)
\end{array}\right. ,\quad 
    \left\{\begin{array}{l}
        w^{3,a}=0   \\
        w^{3,b}=1
    \end{array}\right. .
\end{gather*}
$\mathbf{u}^{2,3}$ is a linear combination of $\mathbf w^2$ and $\mathbf w^3$.
\begin{gather*}
    \mathbf{u}^{2,3}=c_2\mathbf w^2+c_3\mathbf w^3=\begin{pmatrix}
        u_1^{2,3,A} & u_1^{2,3,B} & u_2^{2,3,A} & u_2^{2,3,B} & \cdots & u_N^{2,3,A} & u_N^{2,3,B}
    \end{pmatrix}^\mT ,
\end{gather*}
where
\begin{gather*}
	 \begin{pmatrix}
        u_l^{2,3,a}\\
        u_l^{2,3,b}
    \end{pmatrix}=\beta_2^l\begin{pmatrix}
        l_2+m\beta_2+l_3\beta_2^2\\
        -(l_1+n\beta_2+l_1\beta_2^2)
    \end{pmatrix}c_2+\beta_3^l\begin{pmatrix}
        0\\
        1
    \end{pmatrix}c_3.
\end{gather*}
The boundary condition Eq.~(\ref{Boundary1}) renders
\begin{gather}
    \left(
\begin{array}{cc}
 \beta_2 l_2 n -\beta_2 l_1 (\beta_2 (l_3-l_2)+m) & \beta_3 (\beta_3 l_3+m) \\
 \beta_2 (\beta_2 l_2+m) (l_2+\beta_2 (\beta_2 l_3+m)) & 0 \\
\end{array}
\right)\begin{pmatrix}
    c_2 \\ c_3
\end{pmatrix}=0. \label{phi2phi3boundary1}
\end{gather}
Nontrivial solution to Eq.~(\ref{phi2phi3boundary1}) implies
\begin{gather}
	\det \left(
\begin{array}{cc}
 \beta_2 l_2 n -\beta_2 l_1 (\beta_2 (l_3-l_2)+m) & \beta_3 (\beta_3 l_3+m) \\
 \beta_2 (\beta_2 l_2+m) (l_2+\beta_2 (\beta_2 l_3+m)) & 0 \\
\end{array}
\right)=0.\label{phi2phi3boundary2}
\end{gather}
Eq.~(\ref{phi2phi3boundary2}) together with Eq.~(\ref{beta2}), Eq.~(\ref{beta3}) renders
\begin{gather}
2m l_2\beta_2 +2 l_2 (l_2+l_3)=2m l_3\beta_3 +2 l_2 (l_2+l_3)=0.\label{brahms}
\end{gather}
Eq.~(\ref{brahms}) together with Eq.~(\ref{beta2}) and Eq.~(\ref{beta3})  enforces $|\beta_2|=1$, which contradicts our assumption $|\beta_2|<1,|\beta_3|<1$. 

\

\paragraph*{Case IV}
$|\beta_1|<1$, $|\beta_2|>1$, $|\beta_3|>1$, $|\beta_4|<1$.

Since $|\beta_1|<1,|\beta_4|<1$, $\mathbf u^{1,4}$ is required to satisfy the boundary condition specified by Eq.~(\ref{Boundary1}). 
Denote the boundary bases corresponding to $\beta_1$ and $\beta_4$ as $\mathbf w^1$ and $\mathbf w^4$, respectively.
\begin{gather*}
   \mathbf{w}^j=\begin{pmatrix}
        \beta_j w^{j,a} & \beta_j w^{j,b} & \beta_j^2 w^{j,a} & \beta_j^2 w^{j,b} & \cdots & \beta_j^N w^{j,a} &\beta_j^N w^{j,b}
    \end{pmatrix}^\mT,\quad j=1,4.
\end{gather*}
The bulk recurrence relation Eq.~(\ref{Bulk1}) gives
\begin{gather}
    \begin{pmatrix}
        l_1+n\beta_1+l_1\beta_1^2 & 0 \\
        l_3+m\beta_1+l_2\beta_1^2 & 0 
    \end{pmatrix}\begin{pmatrix}
        w^{1,a}\\ w^{1,b}
    \end{pmatrix}=0,\quad 
    \begin{pmatrix}
        l_1+n\beta_4+l_1\beta_4^2 & l_2+m\beta_4+l_3\beta_4^2 \\
        0 & 0 
    \end{pmatrix}\begin{pmatrix}
        w^{4,a}\\ w^{4,b}
    \end{pmatrix}=0. \label{bruckner}
\end{gather}
Eq.~(\ref{bruckner}) is solved by
\begin{gather*}
    \left\{\begin{array}{l}
        w^{1,a}=0   \\
        w^{1,b}=1
    \end{array}\right., \quad
    \left\{\begin{array}{l}
     w^{4,a}=l_2+m\beta_4+l_3\beta_4^2 \\
     w^{4,b}=-(l_1+n\beta_4+l_1\beta_4^2)
\end{array}\right..
\end{gather*}
$\mathbf{u}^{1,4}$ is a linear combination of $\mathbf w^1$ and $\mathbf w^4$.
\begin{gather*}
    \mathbf{u}^{1,4}=c_1\mathbf w^1+c_4\mathbf w^4=\begin{pmatrix}
        u_1^{1,4,A} & u_1^{1,4,B} & u_2^{1,4,A} & u_2^{1,4,B} & \cdots & u_N^{1,4,A} & u_N^{1,4,B}
    \end{pmatrix}^\mT,
\end{gather*}
where
\begin{gather*}
	\begin{pmatrix}
        u_l^{1,4,A}\\
        u_l^{1,4,B}
    \end{pmatrix}=\beta_1^l\begin{pmatrix}
        0\\
        1
    \end{pmatrix}c_1+\beta_4^l\begin{pmatrix}
        l_2+m\beta_4+l_3\beta_4^2\\
        -(l_1+n\beta_4+l_1\beta_4^2)
    \end{pmatrix}c_4.
\end{gather*}
The boundary condition Eq.~(\ref{Boundary1}) renders
\begin{gather}
    \left(
\begin{array}{cc}
 \beta_1 (\beta_1 l_3+m) & \beta_4 (l_2 n-l_1 (\beta_4 (l_3-l_2)+m)) \\
 0 & \beta_4 (\beta_4 l_2+m) (l_2+\beta_4 (\beta_4 l_3+m)) \\
\end{array}
\right)\begin{pmatrix}
    c_1\\ c_4
\end{pmatrix}=0.\label{chopin}
\end{gather}
Nontrivial solution to Eq.~(\ref{chopin}) requires
\begin{gather}
	\det \left(
\begin{array}{cc}
 \beta_1 (\beta_1 l_3+m) & \beta_4 (l_2 n-l_1 (\beta_4 (l_3-l_2)+m)) \\
 0 & \beta_4 (\beta_4 l_2+m) (l_2+\beta_4 (\beta_4 l_3+m)) \\
\end{array}
\right)=0.\label{argerich}
\end{gather}
Eq.~(\ref{argerich}) along with Eq.~(\ref{beta1}) and Eq.~(\ref{beta4}) renders
\begin{gather}
	2ml_2\beta_4+2l_2(l_2+l_3)=2ml_3\beta_1+2l_2(l_2+l_3)=0.\label{vivaldi}
\end{gather}
Eq.~(\ref{vivaldi}), Eq.~(\ref{beta1}) and Eq.~(\ref{beta4}) enforce $|\beta_4|=1$,
which contradicts our assumption $|\beta_1|<1$, $|\beta_4|<1$.

\subsection{Discussion on the boundary condition Eq.~(\ref{Boundary2})\label{chewinggum2}}

\paragraph*{Case I} $|\beta_1|>1$, $|\beta_2|<1$, $|\beta_3|>1$, $|\beta_4|<1$.

Since $|\beta_1|>1$, $|\beta_3|>1$, $\mathbf u^{1,3}$ is required to satisfy the boundary condition Eq.~(\ref{Boundary2}). 
The boundary bases corresponding to $\beta_1$ and $\beta_3$ are $\mathbf w^1$ and $\mathbf w^3$, respectively.
\begin{gather*}
    \mathbf{w}^j=\begin{pmatrix}
        \beta_j w^{j,a} & \beta_j w^{j,b} & \beta_j^2 w^{j,a} & \beta_j^2 w^{j,b} & \cdots & \beta_j^N w^{j,a} &\beta_j^N w^{j,b}
    \end{pmatrix}^\mT,\quad j=1,3.
\end{gather*}
The bulk recurrence relation  Eq.~(\ref{Bulk1}) gives
\begin{gather}
    \begin{pmatrix}
        l_1+n\beta_{j}+l_1\beta_{j}^2 & 0 \\
        l_3+m\beta_{j}+l_2\beta_{j}^2 & 0
    \end{pmatrix}\begin{pmatrix}
        w^{j,a}\\ w^{j,b}
    \end{pmatrix}=0,\quad j=1,3 \label{phi1phi3bulk}
\end{gather}
 Eq.~(\ref{phi1phi3bulk}) is solved by
\begin{gather*}
    w^{j,a}=0,\quad 
    w^{j,b}=1,\quad j=1,3.
\end{gather*}
$\mathbf{u}^{1,3}$ is a linear combination of $\mathbf w^1$ and $\mathbf{w}^3$.
\begin{gather*}
\mathbf{u}^{1,3}=c_1\mathbf w^1+c_3\mathbf w^3=\begin{pmatrix}
        u^{1,3,a}_1 & u^{1,3,b}_1 & u^{1,3,a}_2 &  u^{1,3,b}_2 & \cdots & u^{1,3,a}_N &  u^{1,3,b}_N
    \end{pmatrix}^\mT,
\end{gather*}
where
\begin{gather*}
	\begin{pmatrix}
        u_l^{1,3,a}\\
        u_l^{1,3,b}
    \end{pmatrix}=\beta_1^l\begin{pmatrix}
        0 \\ 1
\end{pmatrix}c_1+\beta_3^l\begin{pmatrix}
        0 \\ 1
    \end{pmatrix}c_3.
\end{gather*}
The boundary condition Eq.~(\ref{Boundary2}) yields
\begin{gather}
    \begin{pmatrix}
        \beta_1^{N-1} (l_2+\beta_1 m) & \beta_3^{N-1} (l_2+\beta_3 m) \\
        0 & 0
    \end{pmatrix}\begin{pmatrix}
        c_1 \\ c_3
    \end{pmatrix}=0.\label{contigent}
\end{gather}
Eq.~(\ref{contigent}) always has a nontrivial solution
\begin{gather*}
	c_1=\beta_3^{N-1} (l_2+\beta_3 m)c ,\quad c_3=-\beta_1^{N-1} (l_2+\beta_1 m)c,
\end{gather*}
where $c$ is a normalization constant. Therefore, under the condition
\begin{gather}
	\mathbb D_1 : ~ \left\{
	\begin{array}{l}
		l_4=0,\\
		m\neq 0,\quad l_2l_3\neq 0,\\
		|\beta_1|>1, |\beta_2|<1, |\beta_3|>1, |\beta_4|<1,
	\end{array}
	\right.\label{s3}
\end{gather}
there always exists a boundary zero mode satisfying the boundary condition Eq.~(\ref{Boundary2}).

\

 \paragraph*{Case II} $|\beta_1|<1$, $|\beta_2|>1$, $|\beta_3|<1$, $|\beta_4|>1$.
 
Since $|\beta_2|>1$, $|\beta_4|>1$, $\mathbf u^{2,4}$ is required to satisfy the boundary condition Eq.~(\ref{Boundary2}). 
The boundary bases corresponding to $\beta_2$ and $\beta_4$ are $\mathbf w^2$ and $\mathbf w^4$, given by
\begin{gather*}
    \mathbf{w}^j=\begin{pmatrix}
        \beta_j w^{j,a} & \beta_j w^{j,b} & \beta_j^2 w^{j,a} & \beta_j^2 w^{j,b} & \cdots & \beta_j^N w^{j,a} &\beta_j^N w^{j,b}
    \end{pmatrix}^\mT,\quad j=2,4.
\end{gather*}
The bulk recurrence relation Eq.~(\ref{Bulk1}) gives
\begin{gather*}
    \begin{pmatrix}
        l_1+n\beta_{j}+l_1\beta_{j}^2 & l_2+m\beta_{j}+l_3\beta_{j}^2\\
        0 & 0
    \end{pmatrix}\begin{pmatrix}
        w^{j,a}\\ w^{j,b}
    \end{pmatrix}=0,\quad j=2,4,
\end{gather*}
which is solved by
\begin{gather*}
    w^{j,a}=l_2+m\beta_j+l_3\beta_j^2,\quad
    w^{j,b}=-(l_1+n\beta_j+l_1\beta_j^2).
\end{gather*}
$\mathbf{u}^{2,4}$ is a linear combination of $\mathbf w^2$ and $\mathbf w^4$.
\begin{gather*}
    \mathbf{u}^{2,4}=c_2\mathbf w^2+c_4\mathbf w^4=\begin{pmatrix}
        u^{2,4,a}_1 & u^{2,4,b}_1 & u^{2,4,a}_2 &  u^{2,4,b}_2 & \cdots & u^{2,4,a}_N &  u^{2,4,b}_N
    \end{pmatrix}^\mT,
\end{gather*}
where
\begin{gather*}
	\begin{pmatrix}
        u_l^{2,4,a}\\
        u_l^{2,4,b}
    \end{pmatrix}=\beta_2^l\begin{pmatrix}
        l_2+m\beta_2+l_3\beta_2^2 \\ -(l_1+n\beta_2+l_1\beta_2^2)
    \end{pmatrix}c_2+\beta_4^l\begin{pmatrix}
        l_2+m\beta_4+l_3\beta_4^2 \\ -(l_1+n\beta_4+l_1\beta_4^2)
    \end{pmatrix}c_4.
\end{gather*}
The boundary condition Eq.~(\ref{Boundary2}) yields
\begin{gather}
	\left(
\begin{array}{cc}
 -\beta _2^{N+1} \left(l_1 \left(l_2-l_3+\beta _2 m\right)-\beta _2 l_3 n\right) & -\beta _4^{N+1} \left(l_1 \left(l_2-l_3+\beta _4 m\right)-\beta _4 l_3 n\right) \\
 \beta _2^{N-1} \left(l_3+\beta _2 m\right) \left(\beta _2 \left(\beta _2 l_3+m\right)+l_2\right) & \beta _4^{N-1} \left(l_3+\beta _4 m\right) \left(\beta _4 \left(\beta _4 l_3+m\right)+l_2\right) \\
\end{array}
\right)\begin{pmatrix}
	c_2\\ c_4
\end{pmatrix}=0.
\end{gather}
Nontrivial solution to the boundary condition Eq.~(\ref{Boundary2}) renders
\begin{gather*}
     (l_2-l_3) (n (l_2+l_3)-2 l_1 m)=0.
\end{gather*}
$l_2=l_3$ renders $|\beta_2|=|\beta_4|=1$ or $|\beta_2||\beta_4|=1$, which contradicts  the assumption that $|\beta_2|>1,|\beta_4|>1$. 
Therefore, if 
\begin{gather}
\mathbb D_2 : ~ \left\{
\begin{array}{l}
l_4=0,\\
m\neq 0, \quad l_2\neq l_3,\quad l_2l_3\neq 0 \\
|\beta_1|<1, |\beta_2|>1, |\beta_3|<1, |\beta_4|>1,\\
	n(l_2+l_3)=2l_1 m, 	
\end{array}
\right.\label{s4}
\end{gather}
there exists a boundary zero mode $\mathbf u^{2,4}$ satisfying the boundary condition Eq.~(\ref{Boundary2}). 

\ 

\paragraph*{Case III} $|\beta_1|>1$, $|\beta_2|<1$, $|\beta_3|<1$, $|\beta_4|>1$.

Since $|\beta_1|>1,|\beta_4|>1$, $\mathbf u^{1,4}$ is required to satisfy the boundary condition Eq.~(\ref{Boundary1}).
The boundary bases corresponding to $\beta_1$ and $\beta_4$ are denoted as $\mathbf w^1$ and $\mathbf w^4$, respectively.
\begin{gather*}
    \mathbf{w}^j=\begin{pmatrix}
        \beta_j w^{j,a} & \beta_j w^{j,b} & \beta_j^2 w^{j,a} & \beta_j^2 w^{j,b} & \cdots & \beta_j^N w^{j,a} &\beta_j^N w^{j,b}
    \end{pmatrix}^\mT,\quad j=1,4.
\end{gather*}
The bulk recurrence relation Eq.~(\ref{Bulk1}) gives
\begin{gather}
    \begin{pmatrix}
        l_1+n\beta_1+l_1\beta_1^2 & 0 \\
        l_3+m\beta_1+l_2\beta_1^2 & 0 
    \end{pmatrix}\begin{pmatrix}
        w^{1,a}\\ w^{1,b}
    \end{pmatrix}=0,\quad 
    \begin{pmatrix}
        l_1+n\beta_4+l_1\beta_4^2 & l_2+m\beta_4+l_3\beta_4^2 \\
        0 & 0 
    \end{pmatrix}\begin{pmatrix}
        w^{4,a}\\ w^{4,b}
    \end{pmatrix}=0 .   \label{philipglass}
\end{gather}
Eq.~(\ref{philipglass}) is solved by
\begin{gather*}
\left\{\begin{array}{l}
        w^{1,a}=0   \\
        w^{1,b}=1
    \end{array}\right.,\quad
\left\{\begin{array}{l}
     w^{4,a}=l_2+m\beta_4+l_3\beta_4^2 \\
     w^{4,b}=-(l_1+n\beta_4+l_1\beta_4^2)
\end{array}\right..
\end{gather*}
$\mathbf{u}^{1,4}$ is a linear combination of $\mathbf{w}^1$ and $\mathbf{w}^4$.
\begin{gather*}
    \mathbf{u}^{1,4}=c_1\mathbf w^1+c_4\mathbf w^4=\begin{pmatrix}
        u_1^{1,4,a} & u_1^{1,4,b} & u_2^{1,4,a} & u_2^{1,4,b} & \cdots & u_N^{1,4,a} & u_N^{1,4,b}
    \end{pmatrix}^\mT,
\end{gather*}
where
\begin{gather*}
	\begin{pmatrix}
        u_l^{1,4,a}\\
        u_l^{1,4,b}
    \end{pmatrix}=\beta_1^l\begin{pmatrix}
        0\\
        1
    \end{pmatrix}c_1+\beta_4^l\begin{pmatrix}
        l_2+m\beta_4+l_3\beta_4^2\\
        -(l_1+n\beta_4+l_1\beta_4^2)
    \end{pmatrix}c_4.
\end{gather*}
The boundary condition Eq.~(\ref{Boundary2}) renders
\begin{gather}
    \left(
\begin{array}{cc}
 \beta_1^{N-1} (l_2+\beta_1 m) & \beta_4^{N+1} (\beta_4 l_3 n-l_1 (l_2-l_3+\beta_4 m)) \\
 0 & \beta_4^{N-1} (l_3+\beta_4 m) (l_2+\beta_4 (\beta_4 l_3+m)) \\
\end{array}
\right)\begin{pmatrix}
    c_1 \\ c_4
\end{pmatrix}=0.\label{shostakovich}
\end{gather}
Nontrivial solution to Eq.~(\ref{shostakovich}) implies
\begin{gather}
	\det \left(
\begin{array}{cc}
 \beta_1^{N-1} (l_2+\beta_1 m) & \beta_4^{N+1} (\beta_4 l_3 n-l_1 (l_2-l_3+\beta_4 m)) \\
 0 & \beta_4^{N-1} (l_3+\beta_4 m) (l_2+\beta_4 (\beta_4 l_3+m)) \\
\end{array}
\right)=0.\label{busoni}
\end{gather}
Eq.~(\ref{busoni}) together with Eq.~(\ref{beta1}), Eq.~(\ref{beta4}) renders
\begin{gather}
	2ml_2\beta_4+2l_2(l_2+l_3)=2ml_3\beta_1+2l_2(l_2+l_3)=0.\label{sanctimonious}
\end{gather}
Eq.~(\ref{sanctimonious}) together with Eq.~(\ref{beta1}) and Eq.~(\ref{beta4})  enforces $|\beta_4|=1$,
which contradicts our assumption $|\beta_1|>1,|\beta_4|>1$. 

\ 

\paragraph*{Case IV}
$|\beta_1|<1$, $|\beta_2|>1$, $|\beta_3|>1$, $|\beta_4|<1$.

Since $|\beta_2|>1,|\beta_3|>1$, $\mathbf u^{2,3}$ is required to satisfy the boundary condition specified by Eq.~(\ref{Boundary2}).
The boundary bases corresponding to $\beta_2$ and $\beta_3$ are denoted as $\mathbf w^2$ and $\mathbf w^3$, respectively.
\begin{gather*}
    \mathbf{w}^j=\begin{pmatrix}
        \beta_j w^{j,a} & \beta_j w^{j,b} & \beta_j^2 w^{j,a} & \beta_j^2 w^{j,b} & \cdots & \beta_j^N w^{j,a} &\beta_j^N w^{j,b}
    \end{pmatrix}^\mT,\quad j=2,3.
\end{gather*}
The bulk recurrence relation Eq.~(\ref{Bulk1}) yields
\begin{gather}
\begin{pmatrix}
        l_1+n\beta_2+l_1\beta_2^2 & l_2+m\beta_2+l_3\beta_2^2 \\
        0 & 0 
    \end{pmatrix}\begin{pmatrix}
        w^{2,a}\\ w^{2,b}
    \end{pmatrix}=0,\quad 
    \begin{pmatrix}
        l_1+n\beta_3+l_1\beta_3^2 & 0 \\
        l_3+m\beta_3+l_2\beta_3^2 & 0 
    \end{pmatrix}\begin{pmatrix}
        w^{3,a}\\ w^{3,b}
    \end{pmatrix}=0.\label{prokofiev}
\end{gather}
Eq.~(\ref{prokofiev}) is solved by
\begin{gather*}
    \left\{ 
    \begin{array}{l}
         w^{2,a}=l_2+m\beta_2+l_3\beta_2^2\\
    w^{2,b}=-(l_1+n\beta_2+l_1\beta_2^2)
    \end{array}\right.,\quad
    \left\{\begin{array}{l}
          w^{3,a}=0 \\
         w^{3,b}=1
    \end{array}
    \right..
\end{gather*}
$\mathbf{u}^{2,3}$ is a linear combination of $\mathbf{w}^2$ and $\mathbf{w}^3$.
\begin{gather*}
    \mathbf{u}^{2,3}=c_2\mathbf{w}^2+c_3\mathbf{w}^3=\begin{pmatrix}
        u^{2,3,a}_1 & u^{2,3,b}_1 & u^{2,3,a}_2 &u^{2,3,b}_2 &\cdots & u^{2,3,a}_N & u^{2,3,}_N
    \end{pmatrix}^\mT,
\end{gather*}
where
\begin{gather}
	    \begin{pmatrix}
        u_l^{2,3,a}\\
        u_l^{2,3,b}
    \end{pmatrix}=\beta_2^l\begin{pmatrix}
        l_2+m\beta_2+l_3\beta_2^2\\
        -(l_1+n\beta_2+l_1\beta_2^2)
    \end{pmatrix}c_2+\beta_3^l\begin{pmatrix}
        0\\
        1
    \end{pmatrix}c_3
\end{gather}
 The boundary condition (\ref{Boundary2}) renders
\begin{gather}
    \left(
\begin{array}{cc}
 \beta_2^{N+1} (\beta_2 l_3 n-l_1 (l_2-l_3+\beta_2 m)) & \beta_3^{N-1} (l_2+\beta_3 m) \\
 \beta_2^{N-1} (l_3+\beta_2 m) (l_2+\beta_2 (\beta_2 l_3+m)) & 0 \\
\end{array}
\right)\begin{pmatrix}
   c_2\\ c_3
\end{pmatrix}=0\label{berg}
\end{gather}
Nontrivial solution to Eq.~(\ref{berg}) requires
\begin{gather}
	\det \left(
\begin{array}{cc}
 \beta_2^{N+1} (\beta_2 l_3 n-l_1 (l_2-l_3+\beta_2 m)) & \beta_3^{N-1} (l_2+\beta_3 m) \\
 \beta_2^{N-1} (l_3+\beta_2 m) (l_2+\beta_2 (\beta_2 l_3+m)) & 0 \\
\end{array}
\right)=0.\label{horowitz}
\end{gather}
Eq.~(\ref{horowitz}), Eq.~(\ref{beta2}) and Eq.~(\ref{beta3}) yield
\begin{gather}
	2ml_2\beta_2+2l_2(l_2+l_3)=2ml_3\beta_3+2l_2(l_2+l_3)=0.\label{liszt}
\end{gather}
Eq.~(\ref{liszt}), Eq.~(\ref{beta2}) and Eq.~(\ref{beta3}) enforce $|\beta_2|=1$,
which contradicts our assumption $|\beta_2|>1$, $|\beta_3|>1$.

\subsection{The conditions for boundary zero modes under $\det B \neq 0, l_4=0$}

The set operation $\mathbb C_1\cup \mathbb C_2$ 
 yields the condition for the appearance of a boundary zero mode satisfying the boundary condition Eq.~(\ref{Boundary1}) under $\det B\neq 0$ and $l_4=0$:
\begin{gather}
	\mathbb C_1:\ \left\{
\begin{array}{l}
l_4=0,\quad n(l_2+l_3)=2l_1 m,\\
m\neq 0,\quad  l_2l_3\neq 0, \\
|m|<|l_2+l_3|,\quad |l_3|<|l_2|,\\
\end{array}
\right.\quad \text{or} \quad 
\mathbb C_2:\ \left\{\begin{array}{l}
	    l_4=0,\\
		m\neq 0,\quad  l_2l_3\neq 0,\\
		|m|<|l_2+l_3|,\quad |l_2|<|l_3|.
	\end{array}\right.
\end{gather}
Here, the original condition $|\beta_1|>1$, $|\beta_2|<1$, $|\beta_3|>1$, $|\beta_4|<1$ simplifies to 
$$
\left\{\begin{array}
	{l}|m|<|l_2+l_3|, \\ |l_3|<|l_2|,
\end{array}\right.
$$
and the original condition $|\beta_1|<1$, $|\beta_2|>1$, $|\beta_3|<1$, $|\beta_4|>1$ simplifies to
$$
\left\{\begin{array}
	{l}|m|<|l_2+l_3|, \\ |l_2|<|l_3|.
\end{array}\right.
$$
Likewise, the condition for the appearance of a boundary zero mode satisfying the boundary condition Eq.~(\ref{Boundary2}) given by the set operation $\mathbb D_1\cup \mathbb D_2$ reads
\begin{gather}
	\mathbb D_1:\ \left\{\begin{array}{l}
	    l_4=0,\\
		m\neq 0,\quad  l_2l_3\neq 0,\\
		|m|<|l_2+l_3|,\quad |l_3|<|l_2|,
	\end{array}\right. \quad \text{or} \quad 
	\mathbb D_2:\ \left\{
\begin{array}{l}
l_4=0,\quad n(l_2+l_3)=2l_1 m,\\
m\neq 0,\quad  l_2l_3\neq 0, \\
|m|<|l_2+l_3|,\quad |l_2|<|l_3|.\\
\end{array}
\right.
\end{gather}
The condition for the existence of boundary zero modes is denoted as $\mathbb U=\mathbb C_1 \cup \mathbb C_2 \cup \mathbb D_1 \cup \mathbb D_2$. The condition for the appearance of exactly one boundary zero mode is
\begin{gather}
\left\{\begin{array}{l}
	l_4=0,\\
	n (l_2+l_3)\neq 2 l_1 m,\quad l_2\neq l_3,\quad l_2l_3\neq 0,\\
	|m|<|l_2+l_3|,
\end{array}\right.\label{hausdorff1}
\end{gather}
derived from the set operation  $\mathbb  U\backslash [(\mathbb C_1 \cup\mathbb C_2)\cap (\mathbb D_1\cup\mathbb D_2)]$. The zero mode manifests as either:

A lower-boundary zero mode satisfying Eq.~(\ref{Boundary1}) when $|l_2|<|l_3|$, or

An upper-boundary zero mode satisfying Eq.~(\ref{Boundary2}) when $|l_2|>|l_3|$.

The condition for two boundary zero modes is
\begin{gather}
\left\{\begin{array}{l}
	l_4=0,\quad n (l_2+l_3)= 2 l_1 m,\\
	l_2\neq l_3,\quad l_2l_3\neq 0,\\
	|m|<|l_2+l_3|,
\end{array}\right.\label{hausdorff2}
\end{gather}
obtained from the complement of set $\mathbb  U\backslash [(\mathbb C_1 \cup\mathbb C_2)\cap (\mathbb D_1\cup\mathbb D_2)]$, i.e.,  $(\mathbb C_1 \cup\mathbb C_2)\cap (\mathbb D_1\cup\mathbb D_2)$.

\twocolumngrid

\end{document}